\documentclass[prd,preprintnumbers,amsmath,amssymb,floatfix]{revtex4}
\usepackage{graphicx}
\usepackage{bm}
\usepackage{amsfonts}
\usepackage{lineno,hyperref}
\usepackage{array}
\usepackage{color,multirow,cases, makecell}
\usepackage{pifont, float, setspace, moresize}

\begin{document}

\title{Cosmological perturbations and gravitational waves in the general Einstein-vector theory}

\author{Xiao-Bin Lai$^{a}$$^{b}$\footnote{laixb2024@lzu.edu.cn}}
\author{Yu-Zhi Fan$^{a}$$^{b}$\footnote{fanyzh2025@lzu.edu.cn}}
\author{Yu-Qi Dong$^{a}$$^{b}$\footnote{dongyq2023@lzu.edu.cn}}
\author{Yu-Xiao Liu$^{a}$$^{b}$\footnote{liuyx@lzu.edu.cn, corresponding author}}

\affiliation{
$^{a}$Lanzhou Center for Theoretical Physics,
  Key Laboratory of Theoretical Physics of Gansu Province, 
  Key Laboratory of Quantum Theory and Applications of MoE, 
  Gansu Provincial Research Center for Basic Disciplines of Quantum Physics,
  Lanzhou University, Lanzhou 730000, China \\
$^{b}$Institute of Theoretical Physics $\&$ Research Center of Gravitation,
  School of Physical Science and Technology,
  Lanzhou University, Lanzhou 730000, China
}

\begin{abstract}
We investigate the stability and gravitational waves (GWs) in the four-dimensional general Einstein-vector theory on a cosmological background. To study the stability, we systematically perform a Hamiltonian analysis at the linear perturbation level. The stability conditions are easily satisfied for tensor perturbations, but they impose nontrivial constraints on the parameter space for vector and scalar perturbations. In particular, in the presence of a nonzero background vector field, the scalar sector fails to satisfy the stability conditions in the general parameter space. However, imposing the plane-wave condition relaxes these conditions, making them achievable. In the small-scale limit, we further investigate the GW properties of the general Einstein-vector theory within the stable parameter space, including the number of independent modes, their propagation speeds, and observational constraints from GW experiments. We find that there can be at most two tensor modes, two vector modes, and one scalar mode. Notably, without imposing the plane-wave ansatz, no scalar GWs exist within the stable parameter space. Furthermore, vector GWs are forbidden if tensor GWs propagate exactly at the speed of light.
\end{abstract}

\maketitle
\tableofcontents 

\section{Introduction}
The advent of Einstein's general relativity (GR)~\cite{Einstein:1916vd} marked a fundamental milestone in our understanding of gravity. Over the subsequent century, extensive theoretical and observational efforts have led to the development of a wide class of modified gravity theories~\cite{Clifton:2011jh}. These theories are aimed at probing the fundamental nature of gravitational interactions. The first direct detection of gravitational waves (GWs) in 2015~\cite{LIGOScientific:2016aoc,LIGOScientific:2016emj} has further revitalized these efforts, raising long-standing questions concerning the nature of gravity and the theoretical framework that most fundamentally describes it.
\par
General relativity has been extensively tested and validated in both the weak-field and strong-field regimes. In the weak-field limit, classical tests such as the precession of Mercury's perihelion~\cite{einstein1915erklarung}, the deflection of light~\cite{dyson1923determination}, and the Pound-Rebka experiment~\cite{PhysRevLett.3.439,Pound:1960zz} show excellent agreement with its predictions. Strong-field tests, ranging from the orbital decay of the Hulse-Taylor pulsar~\cite{Hulse:1974eb,Taylor:1979zz} to the first direct detection of GWs (GW150914)~\cite{LIGOScientific:2016aoc,LIGOScientific:2016emj} and the imaging of black holes in M87* and Sagittarius A*~\cite{EventHorizonTelescope:2019dse,EventHorizonTelescope:2022wkp,EventHorizonTelescope:2022xqj}, further support the theory. Nevertheless, several fundamental issues remain difficult to address within the framework of GR, including the dark matter problem~\cite{Smith:1936mlg,Zwicky:1937zza}, the dark energy problem~\cite{Peebles:2002gy}, the quantization problem~\cite{tHooft:1974toh,Goroff:1985th}, and the hierarchy problem~\cite{Arkani-Hamed:1998jmv,Randall:1999ee,Randall:1999vf}. These challenges have motivated ongoing efforts to explore modified gravity theories. 
\par
Modified gravity theories can be constructed through various approaches, such as introducing additional fields~\cite{Brans:1961sx,Kostelecky:2003fs}, including higher-order derivatives~\cite{Sotiriou:2008rp,DeFelice:2008wz}, considering extra dimensions~\cite{Kaluza:1921tu,Randall:1999ee}, and modifying the underlying geometric structure~\cite{buchdahl1970non,Heisenberg:2023lru}. Such theories can lead to cosmological and GW phenomenology that differs significantly from that of GR. For example, some theories predict up to six GW polarization modes~\cite{Eardley:1973zuo}, in contrast to the two tensor modes present in GR. Others can account for the accelerated expansion of the universe or the rotation curves of galaxies, providing viable alternatives to dark energy or dark matter~\cite{Nojiri:2006ri,Famaey:2011kh}, respectively. For further related work, see Refs.~\cite{Dong:2026xab,Dong:2025utw,Dong:2025ddi,Fan:2024pex} on GWs, Refs.~\cite{Liu:2025oho,Liu:2025dqg,Chen:2025doc,Zhang:2025jlb,Cao:2024oud} on black holes, Refs.~\cite{Tan:2024qij,E:2025kic,Jia:2024sdk,Jia:2024pdk} on extra dimensions, as well as Refs.~\cite{Tan:2024urn,Tan:2024qgk,Lai:2025nyo,Liu:2024oeq,Jin:2025dvf} on other related aspects. Consequently, stringent theoretical and experimental tests are essential for identifying the framework that offers a more complete description of gravity.
\par
The direct detection of GWs ~\cite{LIGOScientific:2016aoc,LIGOScientific:2016emj} by Advanced LIGO in 2015 marked the dawn of GW astronomy and opened new avenues for probing gravity and the cosmos. Another major milestone in astronomical observations was achieved in 2017 with the first multimessenger detection of a binary neutron star merger, GW170817~\cite{LIGOScientific:2017vwq}, and its electromagnetic counterpart, the gamma-ray burst GRB170817A~\cite{Goldstein:2017mmi}. This event not only placed stringent constraints on the speed of tensor modes, $c_{t}$, namely $-3\times 10^{-15} \le c_{t}-1 \le 7\times 10^{-16}$~\cite{Savchenko:2017ffs}, but also demonstrated the power of multimessenger astronomy. Evidence for a stochastic GW background at nanohertz frequencies has recently emerged from data collected by pulsar timing arrays (PTAs)~\cite{NANOGrav:2023gor,EPTA:2023fyk,Reardon:2023gzh,Xu:2023wog}. This discovery establishes PTAs as a new observational window and a unique probe of GWs in this frequency band. Reference~\cite{Chen:2023uiz} reported a search for an isotropic nontensorial GW background using the 15-year data set from the North American Nanohertz Observatory for GWs, suggesting that scalar transverse correlations may account for the observed stochastic signal. This result strengthens the prospect of detecting additional GW polarization modes through GW observations. To date, the joint LIGO-Virgo-KAGRA network has detected more than three hundred GW events~\cite{LIGO20250826}, providing a wealth of observational data for testing theories of gravity. These advances pave the way toward identifying the most viable theory of gravity among the many alternatives. 
\par
Furthermore, next-generation ground-based GW observatories, including the Einstein Telescope~\cite{Punturo:2010zz} and Cosmic Explorer~\cite{Reitze:2019iox}, are currently under active development. In the context of space-based GW detection, the Laser Interferometer Space Antenna (LISA) mission~\cite{LISA:2017pwj} in Europe is progressing toward construction, while China's Taiji~\cite{Hu:2017mde} and TianQin~\cite{TianQin:2015yph} programs are being rapidly advanced. These forthcoming detectors are expected to play a crucial role in future observational and theoretical studies of GW physics. In particular, LISA is predicted to exhibit significantly enhanced sensitivity to nontensorial GW polarizations in certain frequency regimes~\cite{Tinto:2010hz,LISA:2024hlh}, thereby enabling stringent tests of alternative theories of gravity. It has been shown that, in the high-frequency part of its sensitivity band (above approximately $6\times 10^{-2}$Hz), LISA is more than ten times as sensitive to scalar-longitudinal and vector signals as to tensor and scalar-transverse modes~\cite{Tinto:2010hz}. In the low-frequency part of the band, LISA is expected to be comparably sensitive to tensor and vector modes, while being somewhat less sensitive to scalar modes. Future high-precision measurements of GW polarization modes will provide a powerful tool to test GR and identify the most viable theory of gravity among alternatives.
\par
In this paper, we investigate the stability and GWs in the general Einstein-vector theory on a cosmological background. {In a homogeneous and isotropic cosmological spacetime, scalar, vector, and tensor perturbations decouple under the scalar-vector-tensor (SVT) decomposition. Consequently, these three classes of perturbations can be analyzed independently, which significantly simplifies the subsequent analysis.} We first derive the background equations of motion for the general Einstein-vector theory in the presence of a perfect fluid. By incorporating observational constraints from the present universe, we briefly explore the cosmological implications of the theory, including its background evolution, constraints on the parameter space, and the effective description of dark energy. {We then perform a systematic stability analysis of tensor, vector, and scalar perturbations using the Hamiltonian formalism.} The action is expanded to quadratic order in perturbations, after which the gauge degrees of freedom are fixed and the nondynamical variables are eliminated through the constraint equations. {This procedure yields a physical Hamiltonian and an effective Lagrangian containing only the dynamical degrees of freedom, both of which provide the foundation for our subsequent analyses of stability and GWs. Finally, we investigate the properties of GWs under the corresponding stability conditions,} including the number of independent modes, their propagation speeds, and observational constraints from GW experiments. Since current GW detectors are sensitive primarily to large wavenumbers $|\vec{k}|$, we focus on the small-scale limit ($|\vec{k}| \rightarrow \infty$).
\par
{This paper is organized as follows. In Sec.~\ref{Per-cosBack-1701}, we perform the SVT decomposition of the perturbations, derive the background field equations, and discuss the effective description of dark energy. Section~\ref{TensorPerturbations-section-1808} is devoted to tensor perturbations. We first derive the quadratic action and examine the corresponding stability conditions, and then investigate the properties of tensor GWs in light of observational constraints. In Sec.~\ref{vector-sec.5-1132245}, we derive the physical Hamiltonian and the effective Lagrangian for vector perturbations using the Hamiltonian formalism, constrain the parameter space through stability requirements, and analyze vector GWs in the small-scale limit. In Sec.~\ref{scalar-sec.6-1132247}, we employ the Hamiltonian formalism to derive the stability conditions for scalar perturbations and investigate the propagation properties of scalar GWs in different regions of the parameter space in the small-scale limit. Our conclusions are presented in Sec.~\ref{conclusion-sec.7-1132252}. Finally, the appendices provide a brief introduction to the general Einstein-vector theory (Appendix~\ref{GEV-theory-1171300}), discuss the Schutz-Sorkin action for a perfect fluid (Appendix~\ref{appendix-Schutz-Sorkin-10162254}), and list the explicit forms of several lengthy expressions and coefficients (Appendix~\ref{quantities-2314}).}

\par
Throughout this work, we restrict our analysis to four-dimensional spacetime. Our conventions are as follows: Greek indices ($\mu,\nu,\alpha,\beta,\dots$) label spacetime coordinates, while Latin indices ($i,j,k,\dots$) label spatial coordinates. We adopt the metric signature $(-,+,+,+)$ and work in units where the speed of light is set to unity, $c = 1$.

\section{Perturbations and cosmological background}\label{Per-cosBack-1701}
The general Einstein-vector theory is a vector-tensor theory in general $D$ dimensions, constructed by Lu and Geng in 2016~\cite{Geng:2015kvs}. In addition to the metric $g_{\mu\nu}$, the theory contains a vector field $A^{\mu}$ that couples bilinearly to curvature polynomials of arbitrary order, in such a way that only the Riemann tensor, rather than its derivatives, appears in the equations of motion. The equation of motion for the vector field is linear in $A^{\mu}$ and involves derivatives only up to second order. Consequently, the theory belongs to the class of second-order derivative gravity theories. We briefly introduce the general Einstein-vector theory in Appendix~\ref{GEV-theory-1171300}.
\par
In this paper, we focus on the general Einstein-vector theory (see Appendix~\ref{GEV-theory-1171300} for details) coupled to a perfect fluid described by the Schutz-Sorkin action (see Appendix~\ref{appendix-Schutz-Sorkin-10162254} for details). The action is given by 
\begin{eqnarray}
  S&=&\frac{1}{16\pi G}\int d^{4}x\sqrt{-g} \Big[ R-2\Lambda_{0}-\frac{1}{4}F^2-\frac{\mu_0^2}{2}A^2 +\beta_1R A^2+\beta_2G_{\mu\nu}A^{\mu}A^{\nu}+\beta_{3}E^{(2)}+\beta_{4}E^{(2)}A^{2} \Big] \nonumber
  \\
  &&-\int d^4x \big[ \sqrt{-g}\rho_m(n)+J^{\mu}(\partial_{\mu}\ell+\mathcal{A}_1\partial_{\mu}\mathcal{B}_1+\mathcal{A}_2\partial_{\mu}\mathcal{B}_2) \big].\label{action-all}
\end{eqnarray}
Here, $F_{\mu\nu}=\nabla_{\mu}A_{\nu}-\nabla_{\nu}A_{\mu}$ denotes the field-strength tensor associated with the vector potential $A^{\mu}$, and $F^2=F_{\mu\nu}F^{\mu\nu}$. The parameters $\mu$, $\beta_{1}$, $\beta_{2}$, $\beta_{3}$, $\beta_{4}$ are constants. $G_{\mu\nu}=R_{\mu\nu}-\frac{1}{2}g_{\mu\nu}R$ is the Einstein tensor, and $E^{(2)}=R^{2}-4R^{\mu\nu}R_{\mu\nu}+R^{\mu\nu\alpha\rho}R_{\mu\nu\alpha\rho}$ denotes the Gauss-Bonnet term. The quantity $\rho_{m}$ represents the energy density, $n$ the particle number density, $J^{\mu}$ a vector density, and $\ell$ a scalar. The quantities $\mathcal{A}_1$, $\mathcal{A}_2$, $\mathcal{B}_1$, and $\mathcal{B}_2$ arise from the intrinsic vector perturbations of the matter sector (see Refs.~\cite{DeFelice:2009bx, DeFelice:2016yws}).
\par
Observations indicate that the current universe is highly consistent with a spatially flat geometry~\cite{WMAP:2006bqn}. Accordingly, we will analyze the equations of motion for the general Einstein-vector theory within a spatially flat cosmological background,
\begin{eqnarray}
    ds^2=-dt^2+a^2(t)\delta_{ij}dx^{i}dx^{j}.\label{FRW-metric}
\end{eqnarray}
Here, $a(t)$ is the scale factor. The universe described by this background is spatially homogeneous and isotropic, which correspondingly dictates the choice of the background fields,
\begin{eqnarray}
    &&\bar{A}_{\mu}=(\bar{A}(t),0,0,0),\label{BG-vector}\\
    &&\bar{J}^{\mu}=\big(\bar{J},0,0,0\big).
\end{eqnarray}
The background field $\bar{A}$ is a function of time $t$. Specifically, in a comoving coordinate system, $\bar{J}$ is constant, as given in Eq.~\eqref{matter-partial-t}.

\subsection{Perturbations in the cosmological background}\label{Pertur-FRW-subsec}
{In a spatially homogeneous and isotropic universe, perturbations of fields can always be decomposed into scalar, vector, and tensor components through the SVT decomposition~\cite{Flanagan:2005yc, Jackiw:2003pm}.} Employing the SVT decomposition, the metric $g_{\mu\nu}$, the vector field $A_{\mu}$, the vector density $J^{\mu}$, and the scalar field $\ell$, including their perturbations around the cosmological background, can be written as
\begin{eqnarray}
  &&ds^2=-(1+2\phi_{h})dt^2+2(\lambda_{i}+\partial_{i}\varphi_{h})dx^{i}dt+a^2\big[ \delta_{ij}+h^{\mathrm{TT}}_{ij}+2\partial_{(i}\varepsilon_{j)}+E\delta_{ij}+\partial_{i}\partial_{j}\alpha \big]dx^{i}dx^{j},\label{ds-full-1514}
  \\
  &&A_{\mu}=\bar{A}_{\mu}+(\phi_{a},\; \zeta_{i}+\partial_{i}\varphi_{a}),
  \\
  &&J^{\mu}=\bar{J}^{\mu}+(\phi_{m},\; \chi^{i}+\frac{1}{a^2}\delta^{ij}\partial_{j}\varphi_{m}), \label{J-full-1514}
  \\
  &&\ell=\bar{\ell}(t)+\phi_{\ell}. \label{ell-full-1514}
\end{eqnarray}
Here, $h^{\mathrm{TT}}_{ij}$ is a transverse-traceless spatial tensor, and $\lambda_{i},\varepsilon_{i},\zeta_{i},\chi^{i}$ are transverse spatial vectors, that is, they satisfy
\begin{eqnarray}
  &&\partial^{i}h^{\mathrm{TT}}_{ij}=0,\quad \delta^{ij}h^{\mathrm{TT}}_{ij}=0,
  \\
  &&\partial^{i}\lambda_{i}=0,\quad\partial^{i}\varepsilon_{i}=0,\quad\partial^{i}\zeta_{i}=0,\quad\partial_{i}\chi^{i}=0,
\end{eqnarray}
where $\partial^{i}=\delta^{ij}\partial_{j}$. The background quantity $\bar{\ell}$ depends only on $t$, as will be shown in Eq.~\eqref{ellBk-1614}. All perturbations, the tensor perturbation $(h^{\mathrm{TT}}_{ij})$, the vector perturbations $(\lambda_{i},\varepsilon_{i},\zeta_{i},\chi^{i})$, and the scalar perturbations $(\phi_{h},\varphi_{h},E,\alpha,\phi_{a},\varphi_{a},\phi_{m},\varphi_{m},\phi_{\ell})$, are functions of the coordinates $(t, x, y, z)$. Although $J^{\mu}$ is a vector density, the decomposition in Eq.~\eqref{J-full-1514} remains valid because the first-order perturbation of $\sqrt{-g}$ vanishes and $\sqrt{-\bar{g}}$ is a function of $t$ only.
\par
Specifically, for $\mathcal{A}_1$, $\mathcal{A}_2$, $\mathcal{B}_1$, and $\mathcal{B}_2$, we adopt the simplest choice, which nevertheless retains all the information required to describe the vector perturbations of matter~\cite{DeFelice:2009bx, DeFelice:2016yws}
\begin{eqnarray}
  \mathcal{A}_1=\delta\mathcal{A}_1(t,z),\quad \mathcal{A}_2=\delta\mathcal{A}_2(t,z),\quad \mathcal{B}_1=x+\delta\mathcal{B}_1(t,z),\quad \mathcal{B}_2=y+\delta\mathcal{B}_2(t,z). \label{AB-Perturbations-1117}
\end{eqnarray}
The quantities $\delta\mathcal{A}_1$, $\delta\mathcal{A}_2$, $\delta\mathcal{B}_1$, and $\delta\mathcal{B}_2$ are perturbations that depend on $t$ and $z$. We work in a coordinate system where GWs propagate along the $+z$ direction. It is important to note that $\delta\mathcal{A}_{1,2}$ and $\delta\mathcal{B}_{1,2}$ contribute exclusively to the vector perturbations of matter.
\par
In this theory, the scalar, vector, and tensor perturbations are decoupled from each other in the cosmological background. This allows us to treat them separately, greatly simplifying the subsequent analysis and calculations.

\subsection{Background equations} \label{BEofM}
We begin by considering the matter action in Eq.~\eqref{perfect-fluid-action}. From Eq.~\eqref{vectorDensity-all}, we obtain the background value $\bar{J}^{\mu}$ of the vector density $J^{\mu}$:
\begin{eqnarray}
    \bar{J}^{\mu}=(\bar{n} a^3,0,0,0).\label{JBackground-2124}
\end{eqnarray}
Here, we work in comoving coordinates where $U^{\mu}=(1,0,0,0)$. Varying the action in Eq.~\eqref{action-all} with respect to $J^{\mu}$ yields a constraint on $\bar{\ell}$,
\begin{eqnarray}
  \dot{\bar{\ell}}=-\bar{\rho}_{m,n}. \label{ellBk-1614}
\end{eqnarray}
Here, $\partial_i\bar{\ell}=0$ has been omitted, which implies $\bar{\ell}$ is a function of $t$ only. Hereafter, a dot denotes a time derivative (e.g., $\dot{\bar{n}} = \partial \bar{n}/\partial t$). Particle number conservation follows from varying the action~\eqref{action-all} with respect to $\bar{\ell}$, and is expressed as the continuity equation:
\begin{eqnarray}
    0&=&\partial_{\mu}\bar{J}^{\mu}=\partial_{t}(\bar{n} a^3)=\frac{\partial\bar{\rho}_m}{\partial \bar{n}}\dot{\bar{n}} a^3 +3\bar{n} \frac{\partial\bar{\rho}_m}{\partial \bar{n}} a^2 \dot{a} \nonumber\\
    &=&\dot{\bar{\rho}}_m+3H(\bar{\rho}_m+\bar{p}_m).\label{matter-partial-t}
\end{eqnarray}
Here we use the definitions of the Hubble parameter $H = \dot{a}/a$ {and the pressure $p_m=n\frac{\partial\rho_m}{\partial n}-\rho_m$}, and multiply the right-hand side of the third equality by ${\partial\bar{\rho}_m}/{\partial\bar{n}}$. This operation is valid because the left-hand side of the equation is zero.
\par
Under normal circumstances, the energy density $(\bar{\rho}_m)$ is positive and gives rise to a positive pressure $(\bar{p}_m)$. Equation~\eqref{matter-partial-t} implies that, if the universe were static, i.e., $H=0$, the energy density $\bar{\rho}_m$ would be constant. Observations, however, have shown that the present universe is not only expanding but also accelerating~\cite{SupernovaCosmologyProject:1998vns, SupernovaSearchTeam:1998fmf, Riess:1998dv}. For an expanding universe, one has $H(t_0)>0$ at the present time $t_0$, which implies $\dot{\bar{\rho}}_m|_{t=t_{0}}<0$. Thus, as the universe expands, the energy density of ordinary matter decreases, as expected physically.
\par
To derive the Friedmann equation, we introduce the lapse function $N(t)$ into the cosmological metric~\eqref{FRW-metric}
\begin{eqnarray}
    ds^2=-N^2(t)dt^2+a^2(t)\delta_{ij}dx^{i}dx^{j}.\label{metric-N}
\end{eqnarray}
After varying the action in Eq.~\eqref{action-all}, we set $N=1$. In this background, the Schutz-Sorkin action~\eqref{perfect-fluid-action} reduces to
\begin{eqnarray}
    \bar{S}_m=-\int d^4x a^3(N\bar{\rho}_m+\bar{n}\partial_t \bar{\ell}).
\end{eqnarray}
Next, substituting the background metric~\eqref{metric-N} and the background vector field~\eqref{BG-vector} into the action~\eqref{action-all}, we obtain the background action $\bar{S}$. 
\par
The background equations are obtained by varying the action $\bar{S}$ with respect to $N$, $a$, and $\bar{A}$, and subsequently setting $N=1$, $\dot{N}=0$, and $\ddot{N}=0$,
\begin{eqnarray}
    \bar{\rho}_{m}&=&\frac{1}{16\pi G}\Bigg[ \left( 6 H^2-2\Lambda_{0}-\frac{1}{2}\mu_{0}^2\bar{A}^2 \right) +6\beta_{1}\bar{A}\left(3 \bar{A}H^3-2\dot{\bar{A}}H+2\bar{A}\dot{H}\right) +48\beta_{4}\bar{A}H^2\left(\bar{A}H^2-\dot{\bar{A}}H+A\dot{H}\right)\nonumber\\
    &&\qquad\quad -9\beta_{2}\bar{A}^2H^2 \Bigg],\label{EqN-BG}
    \\
    \bar{p}_{m}&=&\frac{-1}{16\pi G}\Bigg[ \left( 6 H^2+4\dot{H}-2\Lambda_{0}+\frac{1}{2}\mu_{0}^2\bar{A}^2 \right) -2\beta_{1}\left( 3\bar{A}^2H^2+4\bar{A}\dot{\bar{A}}H+2\bar{A}^2\dot{H}+2\bar{A}\ddot{\bar{A}}+2\dot{\bar{A}}^2 \right) \nonumber\\
    && \qquad\quad -\beta_2\bar{A}\left( 3\bar{A}H^2+4\dot{\bar{A}}H+2\bar{A}\dot{H} \right) -16\beta_4H \left( 2\bar{A}\dot{\bar{A}}H^2+\left(\dot{\bar{A}}^2+\bar{A}\ddot{\bar{A}}\right)H +\bar{A}\dot{\bar{A}}\dot{H} \right) \Bigg],\label{Eqa-BG}
    \\
    0&=&\bar{A}\left[ \mu_{0}^2 -12\beta_{1}\left(2 H^2+\dot{H}\right) +6\beta_{2}H^2 -48\beta_{4}H^2\left(H^2+\dot{H}\right) \right].\label{EqA-BG}
\end{eqnarray}
\par
For the  Hubble parameter $H$, we consider only its nontrivial solution $H=H(t)$ in this paper. An expanding universe corresponds to $H(t_0)>0$. An accelerating universe further requires $\frac{\ddot{a}(t_0)}{a(t_0)}=H^2(t_0)+\dot{H}(t_0)>0$, which implies $H^2(t_0)>-\dot{H}(t_0)$. For the background vector field $\bar{A}$, we will consider two cases: $\bar{A}=0$ and $\bar{A}\ne 0$.
\par
We begin by considering \textbf{the case} \bm{$\bar{A}=0$}, which reduces the background equations~\eqref{EqN-BG} and~\eqref{Eqa-BG},
\begin{eqnarray}
    &&\dot{H}=-4\pi G(\bar{\rho}_m+\bar{p}_m),\label{H'Eq-1541}\\
    &&H^2=\frac{8\pi G}{3}\bar{\rho}_m+\frac{\Lambda_0}{3}.\label{HEq-1541}
\end{eqnarray}
This is analogous to Einstein's GR with a cosmological constant. Since the current universe is undergoing accelerated expansion, which requires $\frac{\ddot{a}(t_0)}{a(t_0)}=H^2(t_0)+\dot{H}(t_0)>0$, it follows that
\begin{eqnarray}
    \bar{\rho}_m(t_0)+3\bar{p}_m(t_0)<\frac{\Lambda_0}{4\pi G}.
\end{eqnarray}
With positive energy density $\bar{\rho}_{m}$ and pressure $\bar{p}_{m}$, {it follows that
\begin{eqnarray}
  \Lambda_0>0,\quad \dot{H}<0.
\end{eqnarray}
According to Eqs.~\eqref{H'Eq-1541} and~\eqref{HEq-1541},} we obtain 
\begin{eqnarray}
    \frac{\ddot{a}}{a}=H^2+\dot{H}=-\frac{4\pi G}{3}(\bar{\rho}_{m}+3\bar{p}_{m})+\frac{\Lambda_{0}}{3}.
\end{eqnarray}
While both matter and its associated pressure act to suppress cosmic expansion, the cosmological constant $\Lambda_{0}$ conversely promotes it. This promoting effect is commonly attributed to what is termed dark energy.
\par
We now turn to \textbf{the case} \bm{$\bar{A}\ne 0$}. The system of Eqs.~\eqref{EqN-BG}-\eqref{EqA-BG} allows us to solve for the parameters $\mu_{0}^2$, $\dot{H}(t)$, and $\Lambda_{0}$,
\begin{eqnarray}
    \mu_{0}^2&=&12\beta_{1}\left(2H^2+\dot{H}\right)-6\beta_{2}H^2+48\beta_{4}H^2\left(H^2+\dot{H}\right),\label{mu0-solution-952} \\
    \dot{H}&=&\frac{-1}{1-\bar{A}\left(\left(\beta_{1}+\frac{1}{2}\beta_{2}\right)\bar{A}+8\beta_{4}H\dot{\bar{A}}\right)}\left( 4\pi G(\bar{\rho}_{m}+\bar{p}_{m}) -(\beta_{1}+4\beta_{4}H^2)\left(-H\bar{A}\dot{\bar{A}}+\bar{A}\ddot{\bar{A}}+\dot{\bar{A}}^2\right) -\beta_{2}H\bar{A}\dot{\bar{A}} \right),\label{dotH-solution-952} \\
    \Lambda_{0}&=& 3\left(H^2+\dot{H}\right)+4\pi G(\bar{\rho}_{m}+3\bar{p}_{m}) +\beta_{1}\left(3H^2\bar{A}^2-3\dot{\bar{A}}^2-3\bar{A}\left(H\dot{\bar{A}}+\ddot{\bar{A}}\right)\right) -\frac{3}{2}\beta_{2}\bar{A}\left(2H^2\bar{A}+2H\dot{\bar{A}}+\dot{H}\bar{A}\right) \nonumber\\
    && 12\beta_{4}H\left(\left(H^3+H\dot{H}\right)\bar{A}^2-H\dot{\bar{A}}^2-H\bar{A}\ddot{\bar{A}}-\left(H^2+2\dot{H}\right)\bar{A}\dot{\bar{A}}\right).\label{Lambda-solution-952}
\end{eqnarray}
In the absence of clear evidence for deviations from GR, it is reasonable to assume that $|\beta_{1}|,|\beta_{2}|,|\beta_{4}|\ll 1$. This assumption, combined with Eqs.~\eqref{mu0-solution-952}-\eqref{Lambda-solution-952}, leads to the finding that
\begin{eqnarray}
    \mu_{0}^2 &\ll& 1, \label{mu0-1039}\\
    \dot{H} &\approx& -4\pi G(\bar{\rho}_{m}+\bar{p}_{m}) <0, \label{dotH-1039}\\
    \Lambda_{0} &\approx& \left(3\left(H^2+\dot{H}\right)+4\pi G\left(\bar{\rho}_{m}+3\bar{p}_{m}\right)\right)|_{t=t_{0}} >0. \label{Lambda-1039}
\end{eqnarray}
In Eqs.~\eqref{dotH-1039} and~\eqref{Lambda-1039}, we have imposed positivity of the energy density and pressure. In addition, in Eq.~\eqref{Lambda-1039} we have used the requirement that the present universe is undergoing accelerated expansion. The constraint~\eqref{dotH-1039} is consistent with both GR and cosmological observations~\cite{Camarena:2019moy,Feeney:2017sgx}. Although $\dot{H}$ is rarely discussed directly in cosmology, it can be expressed in terms of the deceleration parameter $q(z)$ as $\dot{H}=-(1+q)H^2$. According to Ref.~\cite{Camarena:2019moy}, the current value of the deceleration parameter is $q_{0}=-0.55$, which implies $\dot{H}=-0.45 H^2$.

\subsection{Dark parts}
Within the general Einstein-vector theory, one can interpret deviations from GR as contributions from dark energy, thereby enabling a framework to analyze it.
\par
We rewrite Eqs.~\eqref{EqN-BG} and~\eqref{Eqa-BG} as
\begin{eqnarray}
    \frac{3}{8\pi G}H^2 &=& \bar{\rho}_m+\bar{\rho}_D,\label{H2-rho-2245}\\
    \frac{1}{4\pi G}\dot{H} &=& -\bar{\rho}_m-\bar{p}_m-\bar{\rho}_D-\bar{p}_D,\label{H'-rho-2245}
\end{eqnarray}
where the specific forms of $\bar{\rho}_D$ and $\bar{p}_D$ are
\begin{eqnarray}
    \bar{\rho}_D &=& \frac{1}{32\pi G}\Bigg[ \left(4\Lambda_0+\mu_0^2\bar{A}^2\right)-12\beta_1\bar{A}\left(3\bar{A}H^2-2\dot{\bar{A}}H+2\bar{A}\dot{H}\right) -96\beta_4\bar{A}H^2\left(\bar{A}H^2-\dot{\bar{A}}H+\bar{A}\dot{H}\right)\nonumber\\
    &&  +18\beta_2\bar{A}^2H^2 \Bigg],\label{rhoD-2032}
    \\
    \bar{p}_D &=& \frac{1}{32\pi G}\Bigg[ \left(-4\Lambda_0+\mu_0^2\bar{A}^2\right)-4\beta_1\left(3\bar{A}^2H^2+4\bar{A}\dot{\bar{A}}H+2\bar{A}^2\dot{H}+2\bar{A}\ddot{\bar{A}}+2\dot{\bar{A}}^2\right)\nonumber\\
    && -2\beta_2\bar{A}\left(3\bar{A}H^2+4\dot{\bar{A}}H+2\bar{A}\dot{H}\right) -32\beta_4H\left( 2\bar{A}\dot{\bar{A}}H^2+\left(\dot{\bar{A}}^2+\bar{A}\ddot{\bar{A}}\right)H+2\bar{A}\dot{\bar{A}}\dot{H} \right) \Bigg].\label{pD-2032}
\end{eqnarray}
Since $\Lambda_0>0$ and, from Eqs.~\eqref{mu0-1039} and~\eqref{Lambda-1039}, $\mu_0^2,|\beta_{1}|,|\beta_{2}|,|\beta_{4}|\ll 1$, these lead to two constraints: $\bar{\rho}_D>0$ and $\bar{p}_D<0$.
\par
According to the specific forms of $\bar{\rho}_D$~\eqref{rhoD-2032} and $\bar{p}_D$~\eqref{pD-2032}, the dark energy equation of state can be written as
\begin{eqnarray}
    w_D&=&\frac{\bar{p}_D}{\bar{\rho}_D}=-1+\frac{\bar{p}_D+\bar{\rho}_D}{\bar{\rho}_D}\nonumber\\
    &=&-1-\frac{2\beta_{1}\left( \dot{\bar{A}}^2+\bar{A}^2\dot{H}+\bar{A}\left(\ddot{\bar{A}}-\dot{\bar{A}}H\right) \right) +\beta_{2}\partial_{t}\left( \bar{A}^2H \right) +8\beta_{4}H\left( \dot{\bar{A}}^2H+\bar{A}\left(\ddot{\bar{A}}H+2\dot{\bar{A}}\dot{H}-\dot{\bar{A}}H^2\right) \right)}{\Lambda_{0}+3\beta_{2}H^2\bar{A}^2 -3\bar{A}\left(\beta_{1}+4\beta_{4}H^2\right)\left(\left(H^2+\dot{H}\right)\bar{A}-2H\dot{\bar{A}}\right)}. \label{wD-state-2002}
\end{eqnarray}
Since $|\beta_{1}|,|\beta_{2}|,|\beta_{4}|\ll 1$, the second term on the right of the final equality vanishes approximately. It follows that the deviation of $w_D$ from $-1$ is determined by $\beta_1$, $\beta_2$, $\beta_4$, and $\bar{A}$. In particular, when $\bar{A}=0$, the equation of state reduces to $w_D=-1$.
\par
Combining Eqs.~\eqref{H2-rho-2245} and~\eqref{H'-rho-2245}, we derive the equation governing the current accelerated expansion of the universe
\begin{eqnarray}
    \frac{\ddot{a}}{a}&=& H^2+\dot{H}=-\frac{4\pi G}{3}(\bar{\rho}_m+3\bar{p}_m+\bar{\rho}_D+3\bar{p}_D).
\end{eqnarray}
Since $\bar{\rho}_m>0$, $\bar{p}_m>0$, and $\bar{\rho}_D>0$, these three terms act to decelerate the expansion. In contrast, only $\bar{p}_D$ can drive acceleration. The observed accelerated expansion of the current universe therefore requires $\bar{p}_D<-\big( \bar{p}_m+\frac{1}{3}(\bar{\rho}_m+\bar{\rho}_D) \big)$.

\section{The tensor perturbations}\label{TensorPerturbations-section-1808}
According to Eq.~\eqref{action-all}, the action of the general Einstein-vector theory with a perfect fluid is a functional of the metric $g_{\mu\nu}$, the vector field $A_{\mu}$, the vector density $J^{\mu}$, and the scalar fields $\ell, \mathcal{A}_{1}, \mathcal{A}_{2}, \mathcal{B}_{1}, \mathcal{B}_{2}$,
\begin{eqnarray}
    S=S\left[g_{\mu\nu}, A_{\mu}, J^{\mu}, \ell, \mathcal{A}_{1}, \mathcal{A}_{2}, \mathcal{B}_{1}, \mathcal{B}_{2}\right].
\end{eqnarray}
Since the equations of motion for the tensor, vector, and scalar perturbations decouple in a cosmological background, they can be analyzed separately. Here, we focus on the tensor perturbations. 
\par
Since the tensor perturbations originate solely from the metric $g_{\mu\nu}$, we write the perturbed line element as
\begin{eqnarray}
    ds^2=-dt^2+a^2(\delta_{ij}+h^{\mathrm{TT}}_{ij})dx^idx^j.
\end{eqnarray}
Here, $h^{\mathrm{TT}}_{ij}$ is a traceless and divergence-free spatial tensor satisfying $\delta^{ij}h^{\mathrm{TT}}_{ij}=0$ and $\partial^ih^{\mathrm{TT}}_{ij}=0$. Choosing the $+z$ direction as the propagation direction of GWs without loss of generality, the nonvanishing components of $h^{\mathrm{TT}}_{ij}$ are
\begin{eqnarray}
    h^{\mathrm{TT}}_{11}=-h^{\mathrm{TT}}_{22}=h_{+}(t,z),\qquad h^{\mathrm{TT}}_{12}=h^{\mathrm{TT}}_{21}=h_{\times}(t,z),
\end{eqnarray}
where $h_{+}(t,z)$ and $h_{\times}(t,z)$ correspond to the two polarization states. Their amplitudes satisfy $|h_{+}| \ll 1$ and $|h_{\times}| \ll 1$.
\par
In the Schutz-Sorkin action~\eqref{perfect-fluid-action}, the terms $J^{\mu}(\partial_{\mu}\ell+\mathcal{A}_1\partial_{\mu}\mathcal{B}_1+\mathcal{A}_2\partial_{\mu}\mathcal{B}_2)$ do not contribute to the tensor perturbations. The perturbative expansions of $\sqrt{-g}$ and $\rho_m(n)$, however, are obtained via standard perturbation methods,
\begin{eqnarray}
    && \sqrt{-g}=a^3 -\frac{a^3}{2}(h_{+}^2 + h_{\times}^2) + \dots ,\\
    && \rho_m(n)=\rho_m(\bar{n}+\delta n)=\bar{\rho}_m + \frac{\bar{n}}{2}\bar{\rho}_{m,n}(h_{+}^2 + h_{\times}^2) + \dots ,
\end{eqnarray}
where $\bar{\rho}_m=\rho_m(\bar{n})$, $\bar{\rho}_{m,n}=\frac{\partial \rho_m}{\partial n}|_{n=\bar{n}}$, and ``$\dots$'' represents the higher-order terms beyond second-order perturbations. Given these relations, the second-order Schutz-Sorkin action for the tensor perturbations takes the form
\begin{eqnarray}
    S^{(2)}_{m|t}=-\int d^4x\frac{a^3}{2}(\bar{n}\bar{\rho}_{m,n}-\bar{\rho}_m)(h_{+}^2 + h_{\times}^2) =-\int d^4x\frac{a^3}{2}\bar{p}_m(h_{+}^2 + h_{\times}^2).
\end{eqnarray}
\par
After expanding the general Einstein-vector action with a perfect fluid~\eqref{action-all} to second order in perturbations, applying the background equation~\eqref{Eqa-BG}, and integrating by parts, we arrive at the total second-order action $S^{(2)}_t=S^{(2)}_{g|t}+S^{(2)}_{m|t}$ in the form
\begin{eqnarray}
    S_t^{(2)}=\int dtd^3x\frac{a^3}{64\pi G}q_t\left[ \left(\dot{h}_{+}^{2}+\dot{h}_{\times}^{2}\right)-c_t^2\bar{g}^{zz}\left((\partial_z h_{+})^2+(\partial_z h_{\times})^2\right) \right],\label{tensor-181001}
\end{eqnarray}
where $q_t$ and $c_t^2$ are given by
\begin{eqnarray}
    && q_t=2-(2\beta_1+\beta_2)\bar{A}^2-16\beta_4\bar{A}\dot{\bar{A}}H,\label{qt}\\
    && c_t^2=\frac{1}{q_t}\left( 2-(2\beta_1-\beta_2)\bar{A}^2-16\beta_4(\dot{\bar{A}}^2+\bar{A}\ddot{\bar{A}}) \right).
\end{eqnarray}
Here, $c_t^2$ denotes the squared propagation speed of the tensor perturbations. The sign of $q_t$ determines whether the kinetic term for $h_b$ ($b=+,\times$) is positive or negative. Thus, to avoid ghost and Laplacian instabilities, we require
\begin{eqnarray}
    && q_t>0,
    \\
    && c_t^2>0.
\end{eqnarray}
The smallness of the parameters ($|\beta_{1}|, |\beta_{2}|, |\beta_{4}| \ll 1$) makes these conditions straightforward to satisfy. Therefore, ghost and Laplacian instabilities are avoided in the tensor sector of the general Einstein-vector theory.
\par
We vary the action $S_t^{(2)}$ with respect to $h_b$ and derive the corresponding tensor perturbation equation
\begin{eqnarray}
    \ddot{h}_b+\Big(3H+\frac{\dot{q}_t}{q_t}\Big)\dot{h}_b-c_t^2\bar{g}^{zz}\partial_z\partial_z h_b=0.\label{motion-hb-2340}
\end{eqnarray}
Compared with the case of GR, the tensor perturbation equation~\eqref{motion-hb-2340} exhibits deviations, including the time dependence of $q_t$ and a deviation of $c_t^2$ from 1. These modifications lead to a difference between the GW speed and the speed of light, as well as to a modified luminosity distance for GWs relative to that of electromagnetic signals~\cite{Belgacem:2017ihm, Amendola:2017ovw, Belgacem:2018lbp}.
\par
A nonzero $\dot{q}_t$ in the friction term in Eq.~\eqref{motion-hb-2340} implies a modified evolution for $h_b$, differing from its behavior in GR,
\begin{eqnarray}
  \dot{q_t}=-2(2\beta_1+\beta_2)\bar{A}\dot{\bar{A}}-16\beta_4(\dot{\bar{A}}^2H+\bar{A}\ddot{\bar{A}}H+\bar{A}\dot{\bar{A}}\dot{H}).\label{qt-151554}
\end{eqnarray}
Clearly, the terms $\beta_1RA^2$, $\beta_2G_{\mu\nu}A^{\mu}A^{\nu}$, and $\beta_4E^{(2)}A^2$ in the action~\eqref{g-action-4d} directly contribute to deviations of the friction term from its counterpart in GR. According to Eq.~\eqref{qt-151554}, if $\bar{A}$ is constant, these deviations vanish. If instead $\bar{A}=\bar{A}(t)$, the deviation disappears only when $\beta_1=\beta_2=\beta_4=0$, in which case the theory reduces to the Einstein-Maxwell theory supplemented by a Gauss-Bonnet term.
\par
All GWs that can be directly detected by current GW detectors have large wavenumbers $|\vec{k}|$ compared to the cosmic scale, where $|\vec{k}|=\sqrt{\vec{k}^2}$ and $\vec{k}^2=\delta^{ij}k_{i}k_{j}$. Therefore, we shall discuss the properties of GWs in the small-scale limit ($|\vec{k}|\rightarrow \infty$). By performing a Fourier expansion of the tensor perturbation $h_{b}$ and substituting it into Eq.~\eqref{motion-hb-2340}, one can straightforwardly derive the dispersion relation for tensor GWs,
\begin{eqnarray}
  w_b^2-c_{t}^2\bar{g}^{zz}k_{z}^2=0,
\end{eqnarray}
where $w_b$ denotes the frequency of tensor GWs, and $c_{t}^2$ can be expressed as
\begin{eqnarray}
  c_{t}^2&=& 1+\frac{2}{q_{t}}\left( \beta_{2}\bar{A}^2+8\beta_{4}(H\bar{A}\dot{\bar{A}}-\dot{\bar{A}}^2-\bar{A}\ddot{\bar{A}}) \right)\nonumber\\
  &=& 1+\beta_{2}\bar{A}^2+8\beta_{4}(H\bar{A}\dot{\bar{A}}-\dot{\bar{A}}^2-\bar{A}\ddot{\bar{A}}) +\mathcal{O}(\beta_{\bullet}^2).
\end{eqnarray}
Here, on the right-hand side of the second equality sign, all contributions of second and higher order in the coefficients $\beta_{1}$, $\beta_{2}$, and $\beta_{4}$ are included in $\mathcal{O}(\beta_{\bullet}^2)$. Obviously, there are two independent tensor modes propagating at the speed $c_{t}$ in the general Einstein-vector theory. From the action~\eqref{action-all}, one can see that the terms $\beta_2G_{\mu\nu}A^{\mu}A^{\nu}$ and $\beta_{4}E^{(2)}A^2$ provide the dominant and direct contributions to deviations of the tensor GW speed from the speed of light, whereas the term $\beta_3E^{(2)}$ does not enter the tensor equation of motion. The term $\beta_1RA^2$ in the action~\eqref{action-all} affects the tensor GW speed only at second order.
\par
On August 17, 2017, a binary neutron star coalescence candidate (GW170817) was observed by Advanced LIGO and Virgo~\cite{LIGOScientific:2017vwq}. Approximately 1.7 seconds later, the Fermi Gamma-ray Burst Monitor independently detected a gamma-ray burst (GRB170817A)~\cite{Savchenko:2017ffs}. The observations placed a tight constraint on the speed of tensor GWs~\cite{LIGOScientific:2017zic, LIGOScientific:2018dkp}, $-3\times 10^{-15}\le c_t-1\le 7\times 10^{-16}$. This bound is so tight that it is widely accepted that tensor GWs propagate at the speed of light. In the general Einstein-vector theory, the condition for the tensor GW speed to be exactly equal to the speed of light is given by
\begin{eqnarray}
    \beta_2\bar{A}^2+8\beta_4(\bar{A} \dot{\bar{A}}H-\dot{\bar{A}}^2-\bar{A}\ddot{\bar{A}})=0. \label{tensorialGW-constraint-1528}
\end{eqnarray}
When $\bar{A}$ is constant, the condition~\eqref{tensorialGW-constraint-1528} requires either $\beta_{2}=0$ or $\bar{A}=0$. For the time-dependent case $\bar{A}=\bar{A}(t)$ with no fine-tuning between functions, the condition~\eqref{tensorialGW-constraint-1528} results in $\beta_{2}=\beta_{4}=0$.
\par
In this section, we have analyzed the dynamics of the tensor perturbations in the general Einstein-vector theory. There are two dynamical degrees of freedom, $h_{+}$ and $h_{\times}$, corresponding to the two tensor modes (see Eq.~\eqref{tensor-181001}). These modes are free of ghost, Laplacian, and tachyonic instabilities under the assumption $|\beta_{1}|,|\beta_{2}|,|\beta_{4}|\ll 1$. We then discussed the properties of tensor GWs in the small-scale limit, finding two propagation modes with the same speed. In light of the stringent constraint from the GW event GW170817 and its electromagnetic counterpart GRB170817A, there is strong justification to assume that tensor GWs propagate at the speed of light. This requirement leads to three viable regions of parameter space: i) $\bar{A}=0$; ii) $\bar{A}=$const., $\beta_{2}=0$; iii) $\beta_{2}=\beta_{4}=0$. These results are summarized in Table~\ref{tensor-conclusion-181017}.
\begin{table}[h]
\centering
\begin{tabular}{|c|c|c|c|l|}
  \hline
  \textbf{Perturbations} & \textbf{d.o.f.} & \textbf{Stability} & \textbf{Number of GW modes} & \textbf{Cases for \bm{$c_{t}=1$}}
  \\ \hline
  Tensor & 2 & $\surd$ & 2 & \makecell[l]{ i) $\bar{A}=0$; \\ ii) $\bar{A}=$const., $\beta_{2}=0$; \\ iii) $\beta_{2}=\beta_{4}=0$.}
  \\ \hline
\end{tabular}
\caption{The dynamics of the tensor perturbations in the general Einstein-vector theory. The conclusions are derived under the assumption $|\beta_{1}|,|\beta_{2}|,|\beta_{4}|\ll 1$. Within this regime, the stability conditions are automatically satisfied, which explains the appearance of the symbol ``$\surd$'' in the table. The number of propagating modes and the GW speed (shown in the penultimate and last columns of the table) are analyzed in the small-scale limit, i.e., $|\vec{k}|\rightarrow\infty$.}
\label{tensor-conclusion-181017}
\end{table}

\section{The vector perturbations}\label{vector-sec.5-1132245}
\subsection{The second-order action of the vector perturbations}
The focus of this section is on the vector perturbations. According to the SVT decomposition, since the full action~\eqref{action-all} is a functional of $g_{\mu\nu}$, $A_{\mu}$, $J^{\mu}$, $\ell$, $\mathcal{A}_1$, $\mathcal{A}_2$, $\mathcal{B}_1$, and $\mathcal{B}_2$, it is straightforward to see that the vector perturbations arise from $g_{\mu\nu}$, $B_{\mu}$, $J^{\mu}$, $\mathcal{A}_1$, $\mathcal{A}_2$, $\mathcal{B}_1$, and $\mathcal{B}_2$. The explicit forms of the perturbed line element, vector field, and vector density are given in Eqs.~(\ref{ds-full-1514})-(\ref{ell-full-1514}),
\begin{eqnarray}
    && ds^2=-dt^2+2\lambda_i dx^idt +a^2\left[\delta_{ij}+(\partial_i\varepsilon_j +\partial_j\varepsilon_i)\right]dx^idx^j, \label{Vds-1726}\\
    && A_{\mu}=(\bar{A},\; \zeta_{i}),\\
    && J^{\mu}=\big(\bar{J},\; \chi^i\big).\label{VJ-1727}
\end{eqnarray}
Here, the perturbations $\lambda_i$, $\varepsilon_i$, $\zeta_{i}$, and $\chi^{i}$ are functions of spacetime coordinates and satisfy the transverse conditions $\partial^{i}\lambda_i=\partial^{i}\varepsilon_i=\partial^{i}\zeta_{i}=\partial_{i}\chi^i=0$. Without loss of generality, we choose the propagation direction of the perturbations to be along the $+z$ axis. Accordingly, $\lambda_i=\lambda_i(t,z)$, $\varepsilon_i=\varepsilon_i(t,z)$, $\zeta_{i}=\zeta_{i}(t,z)$, and $\chi^{i}=\chi^{i}(t,z)$, with $\lambda_z=\varepsilon_z=\zeta_z=\chi^z=0$. The explicit forms of the perturbations for $\mathcal{A}$ and $\mathcal{B}$ are given by Eq.~\eqref{AB-Perturbations-1117}.
\par
{The perturbative expansion of the action~\eqref{action-all} up to second order in the vector perturbations leads to the second-order action with the perfect fluid, given by
\begin{eqnarray}
  S^{(2)}_{v} &=& \frac{1}{16\pi G}\int dtd^3x \Big[ \frac{q_{t}}{4 a}\delta^{pq}\partial_z\lambda_{p}\partial_z\lambda_{q} +\frac{8\pi G\bar{J}\bar{\rho}_{m,n}}{a^2}\delta^{pq}\lambda_{p}\lambda_{q} +\frac{a}{2}\delta^{pq}\dot{\zeta}_{p}\dot{\zeta}_{q}  -\frac{1}{2 a}\delta^{pq}\partial_z\zeta_{p}\partial_z\zeta_{q} \nonumber\\
  && -\frac{a}{2}\left(4\beta_{2}\dot{H}+Q_{\bar{A}}(t)\right)\delta^{pq}\zeta_{p}\zeta_{q} +\frac{8\pi G a^2}{\bar{J}}\bar{\rho}_{m,n}\delta_{pq}\chi^{p}\chi^{q}  -\frac{\beta_{2}\bar{A}}{a}\delta^{pq}\partial_{z}\lambda_{p}\partial_{z}\zeta_{q}  +16\pi G\bar{\rho}_{m,n}\lambda_{p}\chi^{p} \nonumber\\
  && -16\pi G(\chi^{p}+\delta^{pq}\bar{J}\delta\dot{\mathcal{B}}_{q})\delta\mathcal{A}_{p} + \mathcal{L}_{\varepsilon} \Big],\label{vectorActionP-20264292210}
\end{eqnarray}
where
\begin{eqnarray}
  \mathcal{L}_{\varepsilon}=\frac{a^3q_{t}}{4}\delta^{pq}\partial_{z}\dot{\varepsilon}_{p}\partial_{z}\dot{\varepsilon}_{q} -\frac{a}{2}\delta^{pq}\left(2\beta_{2}\bar{A}\zeta_{p}-q_{t}\lambda_{p}\right)\partial_{z}^2\dot{\varepsilon}_{q}.
\end{eqnarray}
Here, indices $p$ and $q$ run over $1$ and $2$, and $Q_{\bar{A}}(t)=\mu_{0}^2-12\beta_{1}\big(2H^2+\dot{H}\big)+6\beta_{2}H^2-48\beta_{4}H^2\big(H^2+\dot{H}\big)$.} We have used the background equations~\eqref{EqN-BG}-\eqref{EqA-BG} and performed integrations by parts. According to the background equation~\eqref{EqA-BG}, it is straightforward to see that $Q_{\bar{A}}(t)$ vanishes when $\bar{A}\neq 0$.

\subsection{Hamiltonian analysis and stability conditions}\label{gaugeSec-vector-2109}
{To analyze the dynamical behavior of the general Einstein-vector theory, all gauge degrees of freedom must first be eliminated. In addition, it is convenient to remove the remaining nondynamical variables within the Hamiltonian formalism \cite{Dirac:1950pj, Dirac:1958sq}, thereby identifying the true propagating degrees of freedom. This procedure allows us to construct a reduced description containing only the physical variables and provides a transparent framework for studying stability and GW properties. In this subsection, we derive the physical Hamiltonian in the reduced phase space and perform a stability analysis. The resulting formulation provides the basis for studying the properties of GWs in the small-scale limit, $|\vec{k}|\to\infty$, in the next subsection.}
\par
Since the general Einstein-vector theory is covariant, the linearized theory is invariant under infinitesimal local coordinate transformations. Let us consider an infinitesimal coordinate transformation that affects the spatial vector sector,
\begin{eqnarray}
    x^{\mu}\rightarrow x^{\mu}+\xi^{\mu},\qquad \xi^{\mu}=(0,\; \xi_{\mathrm{T}}^i),\label{gaugeTransformation-x-1759}
\end{eqnarray}
where $\xi_{\mathrm{T}}^i(x^\mu)$ is a spacetime function with $|\xi_{\mathrm{T}}^i| \ll 1$ and $\partial_i \xi_{\mathrm{T}}^i = 0$. The perturbations of the metric, vector field, and vector density then undergo the corresponding transformations,
\begin{eqnarray}
    \lambda_i &\rightarrow& \lambda_i -a^2\delta_{ik}\dot{\xi}_{\mathrm{T}}^{k},\label{transformations-perturbation1-2124}\\
    \varepsilon_i &\rightarrow& \varepsilon_i -\delta_{ik}\xi_{\mathrm{T}}^k,\\
    \zeta_i &\rightarrow& \zeta_i,\\
    \chi^{i} &\rightarrow& \chi^{i}+\bar{J}\dot{\xi}_{\mathrm{T}}^i.\label{transformations-perturbation4-2124}
\end{eqnarray}
Since the linearized theory is gauge invariant, we can fix the values of certain components in $\lambda_i$, $\varepsilon_i$, $\zeta_i$, and $\chi^i$ using the perturbation transformations~\eqref{transformations-perturbation1-2124}-\eqref{transformations-perturbation4-2124} without affecting the physical results. If one chooses the gauge condition $\lambda_i=0$ or $\chi^{i}=0$, the transformation vector $\xi_{\mathrm{T}}^i$ is not uniquely fixed. Indeed, since $\dot{\xi}_{\mathrm{T}}^i=\dot{\xi}_{f}^i$, where $\xi_{f}^i=\xi_{\mathrm{T}}^i+f_{\mathrm{T}}^i(x,y,z)$, there remains a residual gauge invariance under transformations generated by the vector $f_{\mathrm{T}}^i(x,y,z)$. This indicates that the gauge freedom is not completely fixed. Therefore, we choose the gauge condition,
\begin{eqnarray}
  \varepsilon_i=0.\label{gauge-vector-1635}
\end{eqnarray}
{Adopting this gauge condition, all terms in the action~\eqref{vectorActionP-20264292210} that involve $\varepsilon_i$ vanish, leading to $\mathcal{L}_{\varepsilon}=0$. Consequently, the corresponding Lagrangian in Fourier space is
\begin{eqnarray}
  \mathcal{L}_{v}&=&\left(\frac{q_{t}k_{z}^2}{64\pi G a} +\frac{\bar{J}\bar{\rho}_{m,n}}{2 a^2}\right)\delta^{pq}\lambda_{p}\lambda_{q} +\frac{a}{32\pi G}\delta^{pq}\dot{\zeta}_{p}\dot{\zeta}_{q}  -\left(\frac{k_{z}^2}{32\pi G a}+\frac{a\left(4\beta_{2}\dot{H}+Q_{\bar{A}}(t)\right)}{32\pi G}\right)\delta^{pq}\zeta_{p}\zeta_{q}\nonumber\\
  && +\frac{ a^2}{2\bar{J}}\bar{\rho}_{m,n}\delta_{pq}\chi^{p}\chi^{q}  -\frac{\beta_{2}\bar{A}}{16\pi G a}k_{z}^2\delta^{pq}\lambda_{p}\zeta_{q}  +\bar{\rho}_{m,n}\lambda_{p}\chi^{p} -(\chi^{p}+\delta^{pq}\bar{J}\delta\dot{\mathcal{B}}_{q})\delta\mathcal{A}_{p}.\label{LV-2026572052}
\end{eqnarray}} 
\par
{Starting from the Lagrangian in Eq.~\eqref{LV-2026572052}, we now perform a Hamiltonian analysis. The purpose is to identify the complete set of constraints and eliminate all nondynamical variables, thereby isolating the true dynamical degrees of freedom of the system. To parametrize the phase space, we introduce the following canonical variables:
\begin{eqnarray}
  \{\lambda_{p},P_{\lambda_p}\}=1,\quad \{\zeta_{p},P_{\zeta_{p}}\}=1,\quad \{\chi^{p},P_{\chi^{p}}\}=1,\quad \{\delta\mathcal{A}_{p},P_{\delta\mathcal{A}_{p}}\}=1,\quad \{\delta\mathcal{B}_{p},P_{\delta\mathcal{B}_{p}}\}=1.
\end{eqnarray}
Here, $\{\cdot,\cdot\}$ denotes the Poisson bracket. According to the definition of the conjugate momentum $P_{i}\equiv\partial\mathcal{L}/\partial\dot{Q}_{i}$ together with the Lagrangian~\eqref{LV-2026572052}, we obtain
\begin{eqnarray}
  && P_{\lambda_p}=0,\quad P_{\chi^{p}}=0,\quad P_{\delta\mathcal{A}_{p}}=0,\quad P_{\delta\mathcal{B}_{p}}=-\bar{J}\delta\mathcal{A}_{p},\label{Pdef1-2026571944}\\
  && P_{\zeta_{p}}=\frac{a}{16 \pi G}\dot{\zeta}_{p}.\label{Pdef2-2026571944}
\end{eqnarray}
The relation in Eq.~\eqref{Pdef1-2026571944} represents four primary constraints, indicating that $\lambda_p$, $\chi^{p}$, and $\delta\mathcal{A}_{p}$ are nondynamical variables. From Eq.~\eqref{Pdef2-2026571944}, one can see that the velocity $\dot{\zeta}_p$ appears explicitly, allowing it to be solved in terms of the canonical momentum as
\begin{eqnarray}
  \dot{\zeta}_{p}=\frac{16 \pi G}{a}P_{\zeta_{p}}. \label{Dzeta-2026572002}
\end{eqnarray}
Substituting Eq.~\eqref{Pdef1-2026571944}, the velocity expression~\eqref{Dzeta-2026572002}, and the Lagrangian~\eqref{LV-2026572052} into the definition of the Hamiltonian, $H\equiv\sum_{n}P_{n}\dot{Q}_{n}-\mathcal{L}$, we obtain the corresponding canonical Hamiltonian:
\begin{eqnarray}
  H_{C}^{(v)}&=&\sum_{p}\Bigg( \frac{8\pi G}{a}(P_{\zeta_{p}})^2 -\left( \frac{q_{t}k_{z}^2}{64\pi G a} +\frac{\bar{J}\bar{\rho}_{m,n}}{2 a^2} \right)(\lambda_{p})^2 +\left(\frac{k_{z}^2}{32\pi G a}+\frac{a}{32\pi G}\left(4\beta_{2}\dot{H}+Q_{\bar{A}}(t)\right)\right)(\zeta_{p})^2 \nonumber\\
  && -\frac{ a^2}{2\bar{J}}\bar{\rho}_{m,n}(\chi^{p})^2 +\frac{\beta_{2}\bar{A}}{16\pi G a}k_{z}^2\lambda_{p}\zeta_{p} \Bigg) -\bar{\rho}_{m,n}\lambda_{p}\chi^{p} +\chi^{p}\delta\mathcal{A}_{p}.\label{canonicalH-2026581132}
\end{eqnarray}}
\par
{It is evident from Eq.~\eqref{Pdef1-2026571944} that the system admits four primary constraints,
\begin{eqnarray}
  \tilde{\chi}_{pl}\approx 0,\label{primaryC-202658958}
\end{eqnarray}
where $\tilde{\chi}_{p1}\equiv P_{\lambda_p}$, $\tilde{\chi}_{p2}\equiv P_{\chi^{p}}$, $\tilde{\chi}_{p3}\equiv P_{\delta\mathcal{A}_{p}}$, and $\tilde{\chi}_{p4}\equiv P_{\delta\mathcal{B}_{p}}+\bar{J}\delta\mathcal{A}_{p}$. Here, the symbol ``$\approx$'' denotes weak equality in the sense of Dirac, namely equality on the constraint surface. Taking these primary constraints into account, the total Hamiltonian is given by
\begin{eqnarray}
  H_{T,p}^{(v)}=H_{C}^{(v)}+\sum_{l=1}^{4}u_{l}^{p}\tilde{\chi}_{pl}.
\end{eqnarray}
Here, $u_{l}^{p}$ are Lagrange multipliers enforcing the constraints $\tilde{\chi}_{l}\approx 0$.}
\par
{We now examine the consistency conditions for the primary constraints under time evolution. Requiring the primary constraints to be preserved in time, i.e., $\dot{\tilde{\chi}}_{pl}=\{\tilde{\chi}_{pl},H_{T,p}^{(v)}\}\approx 0$, yields
\begin{eqnarray}
  && \dot{\tilde{\chi}}_{p4}=\overline{J}u_{p3}\approx 0,\quad \dot{\tilde{\chi}}_{p3}=-\overline{J}u_{p4}-\delta_{pq}\chi^{q}\approx 0,\label{primaryT1-202658949}\\
  && \dot{\tilde{\chi}}_{p2}=\bar{\rho}_{m,n}\lambda_{p} +\frac{a^2\bar{\rho}_{m,n}}{\bar{J}}\delta_{pq}\chi^{q} -\delta\mathcal{A}_{p} \approx 0,\label{primaryT2-202658949}\\
  && \dot{\tilde{\chi}}_{p1}=-\frac{\beta_{2}\bar{A}k_{z}^2}{16\pi G a}\zeta_{p} +\bar{\rho}_{m,n}\delta_{pq}\chi^{q} +\left( \frac{q_{t}k_{z}^2}{32\pi G a} +\frac{\bar{J}\bar{\rho}_{m,n}}{a^2} \right)\lambda_{p} \approx 0,\label{primaryT3-202658949}
\end{eqnarray}
where $u_{pl}=\delta_{pq}u_{l}^{q}$. One can readily see that the two relations in Eq.~\eqref{primaryT1-202658949} determine the Lagrange multipliers $u_{3}^{p}$ and $u_{4}^{p}$. In contrast, Eq.~\eqref{primaryT2-202658949} and Eq.~\eqref{primaryT3-202658949} do not determine additional multipliers, but instead generate two secondary constraints:
\begin{eqnarray}
  \tilde{\chi}_{p5}\equiv \dot{\tilde{\chi}}_{p2} \approx 0,\quad \tilde{\chi}_{p6}\equiv \dot{\tilde{\chi}}_{p1} \approx 0.\label{secondaryC-202658959}
\end{eqnarray}
Taking into account the four primary constraints~\eqref{primaryC-202658958} together with the two secondary constraints~\eqref{secondaryC-202658959}, we can express the total Hamiltonian as
\begin{eqnarray}
  H_{T}^{(v)}=H_{C}^{(v)}+\sum_{l=1}^{6}u_{l}^{p}\tilde{\chi}_{pl}.
\end{eqnarray}
Up to this stage, the system contains six constraints in total, namely $\tilde{\chi}_{pl}\approx 0$. The consistency conditions for these constraints further require $\dot{\tilde{\chi}}_{pl}=\{\tilde{\chi}_{pl},H_{T}^{(v)}\}\approx 0$. After explicit calculation, the resulting six consistency equations completely determine the six Lagrange multipliers $u_{l}^{p}$, and no additional constraints emerge. Therefore, the Hamiltonian system possesses exactly six constraints, namely $\tilde{\chi}_{pl}\approx 0$, all of which satisfy the consistency conditions.}
\par
{Now, we examine whether these six constraints are first-class or second-class. This classification \cite{Dirac:1958sq} determines whether the Hamiltonian system contains gauge degrees of freedom and, consequently, the appropriate strategy for reducing the phase space. We first compute the Poisson brackets among the constraints, which define a $6\times 6$ matrix,
\begin{eqnarray}
  \bm{M}_{ls}\equiv \{\tilde{\chi}_{pl},\tilde{\chi}_{ps}\}.
\end{eqnarray}
We then evaluate its determinant, which is found to be nonvanishing,
\begin{eqnarray}
  \text{det }\bm{M}|_{p=1}=\text{det }\bm{M}|_{p=2}=\frac{a^2q_{t}^2\bar{\rho}_{m,n}^2k_{z}^4}{1024\pi^2 G^2}\ne 0.
\end{eqnarray}
This result shows that all six constraints are second-class, implying that the system contains no gauge degrees of freedom.}
\par
{Since all constraints ($\tilde{\chi}_{pl}\approx 0$) are second-class, they can be used to eliminate six phase-space degrees of freedom. Explicitly, the constraints read
\begin{eqnarray}
  && P_{\lambda_p}=0,\quad P_{\chi^{p}}=0,\quad P_{\delta\mathcal{A}_{p}}=0,\label{constraintS1-2026591506}\\
  && \delta\mathcal{A}_{p}=-\frac{1}{\bar{J}}P_{\delta\mathcal{B}_{p}},\quad \lambda_{p}=\frac{2\beta_{2}\bar{A}}{q_{t}}\zeta_{p} +\frac{32\pi G}{a q_{t}k_{z}^2}P_{\delta\mathcal{B}_{p}},\\
  && \delta_{pq}\chi^{q}=-\frac{2\beta_{2}\bar{J}\bar{A}}{a^2q_{t}}\zeta_{p} -\left( \frac{32\pi G\bar{J}}{a^3 q_{t}k_{z}^2}+\frac{1}{a^2\bar{\rho}_{m,n}} \right)P_{\delta\mathcal{B}_{p}}.\label{constraintS3-2026591506}
\end{eqnarray}
These relations show that the auxiliary sector is completely fixed by the constraint structure. In particular, the canonical momenta conjugate to $\lambda_{p}$, $\chi^{p}$, and $\delta\mathcal{A}_{p}$ vanish identically, confirming that these variables do not represent independent dynamical degrees of freedom. Instead, they function as nondynamical auxiliary fields whose values are fully determined by the remaining phase-space variables. For generic $\beta_{2}\bar{A}\neq 0$, both $\lambda_{p}$ and $\chi^{p}$ depend on the canonical variable $\zeta_{p}$ as well as on the momentum $P_{\delta\mathcal{B}_{p}}$, whereas $\delta\mathcal{A}_{p}$ depends only on $P_{\delta\mathcal{B}_{p}}$. In the special case $\beta_{2}\bar{A}=0$, the dependence on $\zeta_{p}$ disappears, and all three variables $\delta\mathcal{A}_{p}$, $\lambda_{p}$, and $\chi^{p}$ are determined solely by $P_{\delta\mathcal{B}_{p}}$. Moreover, since these variables and their conjugate momenta are completely constrained, their equations of motion contain no independent dynamical content. After imposing the second-class constraints, the variables $\lambda_{p}$, $\chi^{p}$, $\delta\mathcal{A}_{p}$, and their conjugate momenta can be eliminated consistently from the Hamiltonian formulation. Consequently, the reduced phase space is parametrized solely by the genuinely dynamical canonical pair associated with the remaining propagating degrees of freedom.}
\par
{Substituting the solutions in Eqs.~\eqref{constraintS1-2026591506}-\eqref{constraintS3-2026591506} into the canonical Hamiltonian~\eqref{canonicalH-2026581132}, we obtain the physical Hamiltonian,
\begin{eqnarray}
  H_{\text{phys}}^{(v)}&=&\sum_{p}\Bigg( \frac{8\pi G}{a}(P_{\zeta_{p}})^2 +\frac{1}{32\pi G}\left( a\left(4\beta_{2}\dot{H}+Q_{\bar{A}}(t)\right) +\frac{k_{z}^2}{a} +\frac{2\beta_{2}^2\bar{A}^2k_{z}^{4}}{32\pi G \bar{J}\bar{\rho}_{m,n}+a q_{t}k_{z}^2} \right)(\zeta_{p})^2 \nonumber \\
  && +F_{1}(t,k_{z})\left( P_{\delta\mathcal{B}_{p}} +\frac{\beta_{2}\bar{A}}{a^2q_{t}F_{1}(t,k_{z})}\zeta_{p} \right)^2 \Bigg).\label{Hphys-2026581733}
\end{eqnarray}
Here, $F_{1}(t,k_{z})=16\pi G/(a^3q_{t}k_{z}^2)+1/(2\bar{J}a^2\bar{\rho}_{m,n})$. The reduced phase space is therefore spanned by the four variables $\{ P_{\zeta_{p}},P_{\delta\mathcal{B}_{p}},\zeta_{p},\delta\mathcal{B}_{p} \}$, while $\delta\mathcal{B}_{p}$ does not appear in the physical Hamiltonian and hence behaves as a free variable at the level of $H_{\text{phys}}^{(v)}$. From the equation of motion $\dot{P}_{\delta\mathcal{B}_{p}}=-\{ P_{\delta\mathcal{B}_{p}}, H_{\text{phys}}^{(v)} \}=0$, it follows that $P_{\delta\mathcal{B}_{p}}$ is a conserved quantity and can be treated as an arbitrary constant of motion. Consequently, the last term in Eq.~\eqref{Hphys-2026581733} acts as an effective independent potential term.}
\par
{We now turn to the stability analysis. Requiring the positivity of the physical energy, we impose $H_{\text{phys}}^{(v)}\ge 0$. From the explicit form of Eq.~\eqref{Hphys-2026581733}, this condition translates into
\begin{eqnarray}
  && \frac{8\pi G}{a} \ge 0, \quad F_{1}(t,k_{z}) \ge 0, \label{stability1-2026581814}\\
  && \frac{1}{32\pi G}\left( a\left(4\beta_{2}\dot{H}+Q_{\bar{A}}(t)\right) +\frac{k_{z}^2}{a} +\frac{2\beta_{2}^2\bar{A}^2k_{z}^{4}}{32\pi G \bar{J}\bar{\rho}_{m,n}+a q_{t}k_{z}^2} \right) \ge 0.\label{stability2-2026581814}
\end{eqnarray}
Given that $G,a,\bar{J},\bar{\rho}_{m,n},q_{t}>0$, the conditions in Eq.~\eqref{stability1-2026581814} are automatically satisfied. The remaining requirement must hold for arbitrary wavenumber $k_{z}$, which allows us to extract the simplified stability criteria
\begin{eqnarray}
  & \beta_{2} \le 0 &\qquad (\bar{A}\ne 0),\label{stabilityV1-2026591528}\\
  & Q_{\bar{A}}(t) +4\beta_{2}\dot{H} \ge 0 &\qquad (\bar{A}=0).\label{stabilityV2-2026591528}
\end{eqnarray}
Here, we have used $\dot{H}<0$ as given in Eq.~\eqref{dotH-1039}. Therefore, the stability properties of the vector sector are strongly controlled by the background value of $\bar{A}$. In the branch $\bar{A}\ne 0$, the absence of pathological modes reduces to the simple sign condition $\beta_{2}\le 0$, indicating that $\beta_{2}$ directly governs the positivity of the physical Hamiltonian. In contrast, for the branch $\bar{A}=0$, the stability criterion is modified. In this case, Eq.~\eqref{stabilityV2-2026591528} ensures that the Hamiltonian remains nonnegative throughout the cosmological evolution. These results highlight that the stability structure of the vector perturbations is highly sensitive to whether the background vector field $\bar{A}$ is switched on or vanishes.
}

\subsection{The small-scale limit and gravitational waves}
{Since currently detectable GWs correspond to modes with sufficiently large wavenumbers compared to the cosmic scale, we focus our analysis on the small-scale limit ($|\vec{k}|\rightarrow \infty$). According to the geodesic deviation equation, GWs are fully characterized by the Riemann tensor. For vector perturbations, the linear perturbation of the Riemann tensor is directly determined by the vector perturbation $\lambda_i$. From the solutions of the constraint equations in Eqs.~\eqref{constraintS1-2026591506}-\eqref{constraintS3-2026591506}, one finds that, when $\beta_{2} \bar{A}=0$, the perturbation $\lambda_i$ depends only on the conjugate momentum of $\delta\mathcal{B}_{p}$. As will be shown below, $\delta\mathcal{B}_{p}$ does not admit nontrivial plane-wave solutions in this case. Consequently, no propagating vector GW modes exist when $\beta_{2} \bar{A}=0$. Therefore, in what follows, we restrict our attention to the branch $\beta_{2} \bar{A}\ne 0$, where propagating vector GW modes may exist.
}
\par
{In the previous subsection, we derived the physical Hamiltonian~\eqref{Hphys-2026581733}, expressed solely in terms of the dynamical canonical variables. Starting from this Hamiltonian, the corresponding effective Lagrangian can be obtained via an inverse Legendre transformation,
\begin{eqnarray}
  \mathcal{L}_{v,\text{eff}} = \frac{a}{32\pi G}\delta^{pq}\dot{\zeta}_{p}\dot{\zeta}_{q} -\frac{2\beta_{2}^2\bar{A}^2+q_{t}}{32\pi G a q_{t}}k_z^2\delta^{pq}\zeta_{p}\zeta_{q} -\frac{4\beta_{2}\dot{H}+Q_{\bar{A}}(t)}{32\pi G}a\delta^{pq}\zeta_{p}\zeta_{q} +\frac{1}{4 F_{1}(t,k_{z})}\left| \dot{\delta\mathcal{B}}_{p} -\frac{2\beta_{2}\bar{A}}{a^2q_{t}}\zeta_{p} \right|^2.\label{effectiveL-20265101452}
\end{eqnarray}
Here, we have used the inverse Legendre transformation $\mathcal{L}_{v,\text{eff}}=\sum_{p}(P_{\zeta_{p}}\dot{\zeta}_{p}+P_{\delta\mathcal{B}_{p}}\dot{\delta\mathcal{B}}_{p})-H_{\text{phys}}^{(v)}$, together with the Hamilton's canonical equations $\dot{\zeta}_{p}=\{\zeta_p,H_{\text{phys}}^{(v)}\}$ and $\dot{\delta\mathcal{B}}_{p}=\{\delta\mathcal{B}_p,H_{\text{phys}}^{(v)}\}$. It is clear that $\delta\mathcal{B}_{p}$ is a cyclic variable, implying the existence of a conserved quantity,
\begin{eqnarray}
  -\frac{1}{2 F_{1}(t,k_{z})}\left( \dot{\delta\mathcal{B}}_{p} -\frac{2\beta_{2}\bar{A}}{a^2q_{t}}\zeta_{p} \right) =C^{(v)}_{p}, \label{conserved-1312}
\end{eqnarray}
where $C^{(v)}_{p}$ is a constant vector. Since we are interested in plane-wave solutions, we set $C^{(v)}_{p}=0$ for simplicity. In a more general setting, $C^{(v)}_p$ can be solved nonlocally in terms of $\zeta_{p}$ (see Ref.~\cite{Dyer:2008hb}), but this possibility is not pursued here. Substituting Eq.~\eqref{conserved-1312} into the effective Lagrangian~\eqref{effectiveL-20265101452} and taking the small-scale limit ($|\vec{k}|\rightarrow \infty$), we obtain a reduced effective Lagrangian involving only the dynamical variable $\zeta_{p}$,
\begin{eqnarray}
  \mathcal{L}_{v,k_{z}\rightarrow\infty} = \frac{a}{32\pi G}\left(\delta^{pq}\dot{\zeta}_{p}\dot{\zeta}_{q} -c^2_{v}\bar{g}^{zz}k_z^2\delta^{pq}\zeta_{p}\zeta_{q}\right),\label{effectiveAction2-vector-1742}
\end{eqnarray}
where $c^2_{v}=1+ 2\beta_{2}^2\bar{A}^2/q_{t}$ denotes the squared propagation speed of the vector modes. It is then evident that the vector sector contains two propagating degrees of freedom, corresponding to $\zeta_{x}$ and $\zeta_{y}$. When $\beta_{2}\bar{A}\ne 0$, once $\zeta_{p}$ is determined, all other variables can be reconstructed from the constraint equations~\eqref{constraintS1-2026591506}-\eqref{constraintS3-2026591506} and \eqref{conserved-1312}. In contrast, when $\beta_{2}\bar{A}= 0$, no nontrivial plane-wave solutions exist for the remaining variables, and hence no vector GWs are present in this branch. Finally,  when $\beta_{2}\bar{A}\ne 0$, the dispersion relation of the vector GWs can be obtained from Eq.~\eqref{effectiveAction2-vector-1742} by varying with respect to $\zeta_{p}$, }
\begin{eqnarray}
  w_{v}^2-c^2_{v}\bar{g}^{33}k_z^2=0,
\end{eqnarray}
where $w_{v}$ denotes the frequency of vector GWs. Thus, when $\beta_{2}\bar{A}\ne 0$, there exist two independent vector GW modes in the general Einstein-vector theory. Since $|\beta_{1}|,|\beta_{2}|,|\beta_{4}|\ll 1$, the propagation speed of vector GWs is slightly greater than $1$, $ c_{v}\approx 1+\beta_{2}^2\bar{A}^2/2 $. It is therefore clear that the dominant contributions to the deviation of the vector GW speed from the speed of light arise from $\beta_{2}$ and the background vector field $\bar{A}$. By contrast, if $\beta_{2}= 0$ or $\bar{A}= 0$, the theory does not admit propagating vector GWs. In this case, the variable $\zeta_{p}$ decouples from gravity at the linear level.
\par
{In this section, we have analyzed the dynamics of vector perturbations in the general Einstein-vector theory under the gauge condition $\varepsilon_i=0$~\eqref{gauge-vector-1635}. The reduced system contains two dynamical variables, $\zeta_{p}$ and $\delta\mathcal{B}_{p}$, as shown in Eq.~\eqref{effectiveL-20265101452}. Regarding stability, the requirement of positive-definite energy leads to the condition $\beta_{2}\le 0$~\eqref{stabilityV1-2026591528} when $\bar{A}\ne 0$, whereas for the branch $\bar{A}= 0$ the corresponding stability criterion is given by Eq.~\eqref{stabilityV2-2026591528}. Concerning GWs, under the small-scale limit and the plane-wave ansatz, the vector sector contains two propagating vector GW modes. In particular, when $\beta_{2}\bar{A}\ne 0$, the propagation speed of the vector GWs exceeds the speed of light. By contrast, if $\beta_{2}= 0$ or $\bar{A} = 0$, the propagation speed of the vector perturbations reduces to the speed of light. However, in this case no propagating vector GW modes exist, and the variable $\zeta_{p}$ decouples from the metric perturbations at the linear level. These results are summarized in Table~\ref{vector-conclusion-181017}.}
\begin{table}[h]
\centering
\begin{tabular}{|c|c|c|c|c|c|}
  \hline
  \textbf{Perturbations} & \textbf{d.o.f.} & \textbf{Case} & \textbf{Stability} & \textbf{Number of GW modes} & \textbf{Speed}
  \\ \hline
  \multirow{3}*{Vector} & \multirow{3}*{4}
      & $\bar{A}=0$ & \eqref{stabilityV2-2026591528} & 0 & -
      \\ \cline{3-6}
      & & $\bar{A}\ne 0,\beta_{2}=0$ & $\beta_{2}\le 0$ & 0 & -
      \\ \cline{3-6}
      & & $\bar{A}\ne 0, \beta_{2}\ne 0$ & $\beta_{2}\le 0$ & 2 & $>1$
  \\ \hline
\end{tabular}
\caption{{Dynamics of vector perturbations in the general Einstein-vector theory. The results are derived under the assumption $|\beta_{1}|,|\beta_{2}|,|\beta_{4}|\ll 1$. The vector sector contains two dynamical variables, $\zeta_{p}$ and $\delta\mathcal{B}_{p}$, corresponding to four degrees of freedom. The column labeled ``Stability" lists the corresponding stability conditions. When $\bar{A} = 0$ or $\beta_{2}= 0$, no propagating vector GW modes exist. The GW modes and their propagation speeds are analyzed in the small-scale limit under the plane-wave ansatz. The symbol ``-'' indicates that the GW propagation speed is not defined, because GWs are absent in these two cases.}}
\label{vector-conclusion-181017}
\end{table}

\section{The scalar perturbations}\label{scalar-sec.6-1132247}
\subsection{The second-order action of the scalar perturbations}
Having separately analyzed the tensor and vector perturbations of the general Einstein-vector theory, we now turn to the scalar sector, focusing on its dynamical properties, the parameter constraints imposed by stability, and the behavior of GWs in the small-scale limit.
\par
The full action~\eqref{action-all} is a functional of $g_{\mu\nu}$, $A_{\mu}$, $J^{\mu}$, $\ell$, $\mathcal{A}_1$, $\mathcal{A}_2$, $\mathcal{B}_1$, and $\mathcal{B}_2$. Since $\mathcal{A}_1$, $\mathcal{A}_2$, $\mathcal{B}_1$, and $\mathcal{B}_2$ contribute only to the vector perturbations of matter, it follows that $g_{\mu\nu}$, $B_{\mu}$, $J^{\mu}$, and $\ell$ give rise to the scalar perturbations. The explicit forms of these perturbations are expressed as (see Eqs.~(\ref{ds-full-1514})-(\ref{ell-full-1514}))
\begin{eqnarray}
  ds^2&=&-(1+2\phi_{h})dt^2+2\partial_{i}\varphi_{h}dx^{i}dt+a^2\left[ \delta_{ij}+E\delta_{ij}+\partial_{i}\partial_{j}\alpha \right]dx^{i}dx^{j},
  \\
  A_{\mu}&=&\bar{A}_{\mu}+\left(\phi_{a},\; \partial_{i}\varphi_{a}\right),
  \\
  J^{\mu}&=&\bar{J}^{\mu}+\left(\phi_{m},\; \frac{1}{a^2}\partial^{i}\varphi_{m}\right), 
  \\
  \ell&=&\bar{\ell}+\phi_{\ell}. 
\end{eqnarray}
Here, there are nine scalar perturbations $(\phi_{h}, \varphi_{h}, E, \alpha, \phi_{a}, \varphi_{a}, \phi_{m}, \varphi_{m}, \phi_{\ell})$, which are functions of spacetime coordinates. Substituting these perturbations into the full action~\eqref{action-all}, performing integrations by parts, and using the background equations~\eqref{EqN-BG}-\eqref{EqA-BG}, we obtain the second-order perturbation action in Fourier space
\begin{eqnarray}
  S^{(2)}_{s}&=&\int dtd^3x \Bigg[ \left(Q_{1}+\bar{A}Q_{2}\vec{k}^2\right)\phi_{h}^2 +\frac{a\vec{k}^2+a^3Q_{\bar{A}}}{32\pi G}\phi_{a}^2-\frac{\bar{\rho}_{m,nn}}{2a^3}\phi_{m}^2 +\frac{\bar{\rho}_{m,n}}{2\bar{J}a^2}\vec{k}^2\varphi_{m}^2 -3a^2H Q_{2}\phi_{h}\dot{\phi}_{a} \nonumber
  \\
  && -\left(Q_{2}\vec{k}^2+Q_{5}\right)\phi_{h}\phi_{a} -\bar{\rho}_{m,n}\phi_{h}\phi_{m} +\dot{\phi}_{m}\phi_{\ell} -\frac{1}{a^2}\vec{k}^2\varphi_{m}\phi_{\ell} +\mathcal{L}_{\alpha} +\mathcal{L}_{E} +\mathcal{L}_{\varphi_{a}} +\mathcal{L}_{\varphi_h} \Bigg]. \label{action-second-1030}
\end{eqnarray}
This action represents the gauge-ready form of the second-order perturbation action, corresponding to the gauge choices in Eq.~\eqref{gauge-3kind-2057}. The specific terms $\mathcal{L}_{\alpha}$, $\mathcal{L}_{E}$, $\mathcal{L}_{\varphi_{a}}$, and $\mathcal{L}_{\varphi_{h}}$ are as follows,
\begin{eqnarray}
  \mathcal{L}_{\alpha}&=& -\frac{\bar{J}}{2}\Big( \bar{\rho}_{m,n}\phi_{h} +\frac{1}{4 a^3}\bar{\rho}_{m,nn}\big( \bar{J}\vec{k}^2\alpha-6\bar{J}E+4\phi_{m} \big) \Big)\vec{k}^2\alpha +\frac{a^2}{2}\Big( Q_{7}\phi_{h} +\bar{A}Q_{2}\dot{\phi}_{h} -Q_{6}\phi_{a} \nonumber\\
  && -Q_{2}\dot{\phi}_{a} +\frac{a q_{t}}{16\pi G}\dot{E} \Big)\vec{k}^2\dot{\alpha},
  \\
  \mathcal{L}_{E}&=& \Big( -\frac{9\bar{J}^2}{8 a^3}\bar{\rho}_{m,nn}+\frac{a c_{t}^2 q_{t}}{64\pi G}\vec{k}^2 \Big)E^2 -\frac{3 a^3q_{t}}{64\pi G}\dot{E}^2 +\frac{a}{2}\Big( 3a Q_{2}\big( \dot{\phi}_{a}-\bar{A}\dot{\phi}_{h} \big) +\frac{\beta_{2}\bar{A}}{4\pi G}\vec{k}^2\varphi_{a} \Big)\dot{E} \nonumber\\
  && +\Bigg[ \left( Q_{8}+\frac{a}{16\pi G}\Big(q_{t}+4\bar{A}^2\big(\beta_{1}+4\beta_{4}(H^2+\dot{H})\big)\Big)\vec{k}^2 \right)\phi_{h} +Q_{9}\dot{\phi}_{h} +Q_{10}\vec{k}^2\varphi_{h} \nonumber\\
  && +\frac{a q_{t}}{16\pi G}\vec{k}^2\dot{\varphi}_{h} +\frac{1}{2 H}\Big( \frac{\bar{A}}{16\pi G}\partial_{t}(a^3Q_{\bar{A}})-\dot{Q}_{5}+(Q_{6}-\dot{Q}_{2})\vec{k}^2 \Big)\phi_{a} -\frac{1}{2 H}\big(Q_{5}+3a^2\dot{H}Q_{2}\big)\dot{\phi}_{a} \nonumber\\
  && +\frac{\bar{A}a}{32\pi G H}(m_{v}^2-4\beta_{2}\dot{H})\vec{k}^2\varphi_{a} +\frac{3\bar{J}\bar{\rho}_{m,nn}}{2 a^3}\phi_{m} \Bigg]E,
  \\
  \mathcal{L}_{\varphi_{a}}&=& \frac{a}{32\pi G}\vec{k}^2\dot{\varphi}_{a}^2 -\frac{a m_{v}^2}{32\pi G}\vec{k}^2\varphi_{a}^2 -\frac{a}{16\pi G}\vec{k}^2\phi_{a}\dot{\varphi}_{a} -\frac{\beta_{2}\bar{A}a H}{4\pi G}\vec{k}^2\phi_{h}\varphi_{a},
  \\
  \mathcal{L}_{\varphi_{h}}&=& \frac{\bar{J}\bar{\rho}_{m,n}}{2 a^2}\vec{k}^2\varphi_{h}^2 +\bar{A}Q_{2}\vec{k}^2\phi_{h}\dot{\varphi}_{h} +\Big( -Q_{4}\phi_{h} +Q_{6}\phi_{a} +Q_{2}\dot{\phi}_{a} +\frac{\bar{\rho}_{m,n}}{a^2}\varphi_{m} \Big)\vec{k}^2\varphi_{h}.
\end{eqnarray}
Here, $\vec{k}^2=\delta^{ij}k_{i}k_{j}$. The quantities $Q_{\bullet}$ are given in Appendix~\ref{quantities-2314}.
\par
The general Einstein-vector theory is covariant, so its linearized version possesses gauge freedom under infinitesimal local coordinate transformations. Analyzing the physical dynamics requires that this freedom is eliminated. We begin by examining an infinitesimal transformation that acts on the scalar sector,
\begin{eqnarray}
  x^{\mu} \rightarrow x^{\mu}+\xi^{\mu},\qquad \xi^{\mu}=(\xi^t,\;\partial^{i}C). \label{transformation-scalar-2001}
\end{eqnarray}
Here, $\xi^t(x^{\mu})$ and $C(x^{\mu})$ are arbitrary spacetime functions satisfying $|\xi^t|\ll 1$ and $|C|\ll 1$. Under this infinitesimal transformation, the perturbation variables $(\phi_{h}, \varphi_{h}, E, \alpha, \phi_{a}, \varphi_{a}, \phi_{m}, \varphi_{m}, \phi_{\ell})$ transform as follows:
\begin{subequations}\label{transformation-scalars-2047}
\begin{align}
    & \phi_{h}\rightarrow \phi_{h}-\dot{\xi}^{t}, \quad \varphi_{h}\rightarrow \varphi_{h}+\xi^{t}-a^2\dot{C}, \quad E\rightarrow E-2\dot{H}\xi^{t}, \quad \alpha\rightarrow \alpha-2C  ,
    \\
    & \phi_{a}\rightarrow \phi_{a}-\dot{\bar{A}}\xi^{t}-\bar{A}\dot{\xi}^t, \quad \varphi_{a}\rightarrow \varphi_{a}-\bar{A}\xi^t ,
    \\
    & \phi_{m}\rightarrow \phi_{m}-\bar{J}\partial^2C, \quad \varphi_{m}\rightarrow \varphi_{m}+a^2\bar{J}\dot{C} .
\end{align}
\end{subequations}
According to the transformation~\eqref{transformation-scalar-2001}, there are two gauge degrees of freedom for the scalar perturbations in the general Einstein-vector theory. Since $\xi^t$ and $C$ are arbitrary functions of spacetime coordinates, one can always choose them appropriately so as to fix the values of certain perturbation variables via the transformation~\eqref{transformation-scalars-2047}, without affecting the physical results. A convenient gauge choice is to set some scalar perturbations to zero. As in Sec.~\ref{gaugeSec-vector-2109}, to fully fix the gauge freedom, we have the following three types of gauge conditions:
\begin{subequations}\label{gauge-3kind-2057}
\begin{align}
  &\text{Gauge I:}\quad\alpha=0,\; E=0.\\
  &\text{Gauge II:}\quad\alpha=0,\; \varphi_{h}=0.\\
  &\text{Gauge III:}\quad\alpha=0,\; \varphi_{a}=0.
\end{align}
\end{subequations}
Next, we derive the stability conditions and analyze the GW characteristics of the general Einstein-vector theory within the constrained parameter space, adopting the gauge conditions specified above.

\subsection{The small-scale limit and stability conditions}\label{gauge-1-1701}
{Gauge degrees of freedom do not affect physical observables. We therefore fix the gauge by setting $\alpha = 0$ and $E = 0$. In this subsection, we employ a Hamiltonian analysis \cite{Dirac:1950pj, Dirac:1958sq} to derive the effective Lagrangian in the small-scale limit, establish the corresponding stability conditions, and identify the viable region of parameter space.}
\par
{Starting from the action in Eq.~\eqref{action-second-1030}, we perform a Hamiltonian analysis in order to identify the complete set of constraints and eliminate all nondynamical variables. To parameterize the phase space, we introduce the following canonical pairs:
\begin{eqnarray}
  && \{ \phi_{h},P_{\phi_{h}} \}=1,\quad \{ \varphi_{h},P_{\varphi_{h}} \}=1,\quad \{ \phi_{a},P_{\phi_{a}} \}=1,\quad \{ \varphi_{a},P_{\varphi_{a}} \}=1,\\
  && \{ \phi_{m},P_{\phi_{m}} \}=1,\quad \{ \varphi_{m},P_{\varphi_{m}} \}=1,\quad \{ \phi_{\ell},P_{\phi_{\ell}} \}=1.
\end{eqnarray}
Using the definition of the conjugate momentum, $P_{i}\equiv\partial\mathcal{L}/\partial\dot{Q}_{i}$, together with the Hamiltonian definition, $H\equiv\sum_{n}P_{n}\dot{Q}_{n}-\mathcal{L}$, we obtain the canonical Hamiltonian associated with the action~\eqref{action-second-1030},
\begin{eqnarray}
  H_{C}^{(s)} = H_{C}^{(s)}\left[P_{\varphi_{a}},\phi_{h},\varphi_{h},\phi_{a},\varphi_{a},\phi_{m},\varphi_{m},\phi_{\ell}\right].\label{HCs-20265111041}
\end{eqnarray}
The explicit expression for $H_{C}^{(s)}$ is presented in Eq.~\eqref{SHCs-20265111043} of Appendix~\ref{quantities-2314}. Following the same Hamiltonian procedure as that used for the vector perturbations in subsection~\ref{gaugeSec-vector-2109}, we derive the complete set of constraints,
\begin{eqnarray}
  && P_{\phi_{a}}\approx 0,\quad P_{\varphi_{m}}\approx 0,\quad P_{\phi_{\ell}}\approx 0,\quad P_{\phi_{m}}-\phi_{\ell}\approx 0,\label{constraintS1-20265111307}\\
  && 4\pi G P_{\phi_{h}} -3 a^3H\bar{A}(\beta_{1}+4\beta_{4}H^2)\phi_{a}\approx 0,\quad 4\pi G P_{\varphi_{h}} -a \bar{A}(\beta_{1}+4\beta_{4}H^2)\vec{k}^2\left( \bar{A}\phi_{h}-\phi_{a} \right)\approx 0, \label{constraintS2-20265111307}\\
  && \bar{J}\bar{\rho}_{m,n}\varphi_{h} +\bar{\rho}_{m,n}\varphi_{m} -\bar{J}\phi_{\ell}\approx 0,\quad K_{1}\left[ P_{\varphi_{a}},\phi_{h},\varphi_{h},\phi_{a},\varphi_{a},\phi_{m},\varphi_{m} \right]\approx 0,\label{constraintS3-20265111307}
\end{eqnarray}
where the explicit form of $K_{1}\left[ P_{\varphi_{a}},\phi_{h},\varphi_{h},\phi_{a},\varphi_{a},\phi_{m},\varphi_{m} \right]$ is given in Eq.~\eqref{K1-20265111308} of Appendix~\ref{quantities-2314}. To determine the nature of these constraints, we compute the determinant of the matrix formed by their Poisson brackets. Since the determinant is nonvanishing, all constraints are identified as second-class constraints \cite{Dirac:1958sq}. Consequently, eight phase-space degrees of freedom can be eliminated, leaving only the physical dynamical variables in the scalar sector.}
\par
{For the case $\bar{A}\ne 0$ and $\beta_{1}+4\beta_{4}H^2\ne 0$, solving the eight constraint equations and substituting the resulting expressions into the canonical Hamiltonian yields a reduced Hamiltonian containing only the dynamical variables, $H_{\text{phys}}^{(s)}=H_{\text{phys}}^{(s)}\left[ P_{\varphi_{h}},P_{\varphi_{a}},P_{\phi_{m}},\varphi_{h},\varphi_{a},\phi_{m} \right]$. In this reduced phase space, the corresponding effective Lagrangian $\mathcal{L}_{s,\text{eff}}=\mathcal{L}_{s,\text{eff}}\left[ \varphi_{h},\varphi_{a},\phi_{m} \right]$ can be obtained through an inverse Legendre transformation. At this stage, the kinetic terms of the three variables remain mixed. To diagonalize the kinetic structure, we introduce the following field redefinitions:
\begin{eqnarray}
  \phi_{1} &\equiv& \varphi_{a}+\frac{F_{4}(t,\vec{k})}{F_{2}(t,\vec{k})}\varphi_{h}+\frac{F_{5}(t,\vec{k})}{F_{2}(t,\vec{k})}\phi_{m}, \label{phi1-20265111931}\\
  \phi_{2} &\equiv& \varphi_{h}+\frac{F_{6}(t,\vec{k})}{F_{3}(t,\vec{k})}\phi_{m}.\label{phi2-20265111931}
\end{eqnarray}
Here, the explicit forms of $F_{\bullet}(t,\vec{k})$ are given in Eqs.~\eqref{F1-20265111942}-\eqref{F5-20265111942} of the Appendix~\ref{quantities-2314}. By substituting $\phi_{1}$ and $\phi_{2}$ into the effective Lagrangian $\mathcal{L}_{s,\text{eff}}$, one can eliminate $\varphi_{a}$ and $\varphi_{h}$. The resulting Lagrangian in the small-scale limit becomes
\begin{eqnarray}
  \mathcal{L}_{s1,\vec{k}\rightarrow\infty}&=& \frac{3a^3H^2q_{t}}{16\pi G}\big(\dot{\phi}_{2}\big)^2 -\frac{F_{7}(t)}{16\pi G}\vec{k}^2\phi_{2}^2 +\frac{a^2\bar{\rho}_{m,n}}{2\bar{J}}\bigg(\frac{\dot{\phi}_{m}}{|\vec{k}|}\bigg)^2  \nonumber\\
  && -\frac{\bar{p}_{m,n}}{2 \bar{J}}\vec{k}^2\bigg(\frac{\phi_{m}}{|\vec{k}|}\bigg)^2 -\frac{1}{16\pi G}\frac{a(\beta_{1}+4\beta_{4}H^2)}{1-2\beta_{1}-8\beta_{4}H^2}\big(|\vec{k}|\dot{\phi}_{1}\big)^2 \nonumber\\
  &&  +\frac{1}{16\pi G}\frac{aH\big( q_{t} -4\bar{A}^2\big((1-2\beta_{2})(\beta_{1}+4\beta_{4}H^2)+4\beta_{4}\dot{H}\big) \big)}{(1-2\beta_{1}-8\beta_{4}H^2)\bar{A}} |\vec{k}|\phi_{2}\big(|\vec{k}|\dot{\phi}_{1}\big),\label{Lsk-20265112138}
\end{eqnarray}
where the explicit expression for $F_{7}(t)$ is provided in Eq.~\eqref{F8-20265112141} of the Appendix~\ref{quantities-2314}. Introducing the conjugate variables $\{\phi_{1},P_{\phi_{1}}\}=1$, $\{\phi_{2},P_{\phi_{2}}\}=1$, and $\{\phi_{m},P_{\phi_{m}}\}=1$, we obtain the corresponding Hamiltonian,
\begin{eqnarray}
  H_{\vec{k}\rightarrow\infty}^{(s1)}&=&\frac{4\pi G}{3a^3H^2q_{t}}\left( P_{\phi_{2}} \right)^2 +\frac{F_{7}(t)}{16\pi G}\vec{k}^2(\phi_{2})^2 +\frac{\bar{J}}{2 a^2\bar{\rho}_{m,n}}\vec{k}^2\left( P_{\phi_{m}} \right)^2 +\frac{\bar{p}_{m,n}}{2 \bar{J}}\left( \phi_{m} \right)^2\nonumber\\
  && -\frac{4\pi G(1-2\beta_{1}-8\beta_{4}H^2)}{a(\beta_{1}+4\beta_{4}H^2)\vec{k}^2}\left( P_{\phi_{1}} -\frac{aH\big( q_{t} -4\bar{A}^2\big((1-2\beta_{2})(\beta_{1}+4\beta_{4}H^2)+4\beta_{4}\dot{H}\big) \big)}{16\pi G(1-2\beta_{1}-8\beta_{4}H^2)\bar{A}}\vec{k}^2\phi_{2} \right)^2.\label{Hs1-20265131119}
\end{eqnarray}
Stability requires that the Hamiltonian be bounded from below, namely that $H_{\vec{k}\rightarrow\infty}^{(s)}\ge 0$. Assuming $|\beta_{1}|,|\beta_{2}|,|\beta_{4}|\ll 1$ and given that $G,a,\bar{J},\bar{\rho}_{m,n},q_{t}>0$, the stability conditions can be written as
\begin{eqnarray}
  && \beta_{1}=\beta_{4}=0,\label{stabilityS1-20265121600}\\
  && \bar{p}_{m,n}\ge 0.\label{stabilityS2-20265121600}
\end{eqnarray}
Condition~\eqref{stabilityS2-20265121600} follows from the requirement that the coefficient of the $\left( \phi_{m} \right)^2$ term be positive, which guarantees the stability of matter perturbations. On the other hand, positivity of the $(\phi_{2})^2$ term requires $\beta_{1}+4\beta_{4}H^2\ge 0$, whereas positivity of the $\left( P_{\phi_{1}} -\cdots \right)^2$ term requires $\beta_{1}+4\beta_{4}H^2\le 0$. Since $H$ is generally time dependent, these two conditions are compatible only when $\beta_{1}=\beta_{4}=0$, which leads directly to the stability condition~\eqref{stabilityS1-20265121600}. However, Eqs.~\eqref{constraintS1-20265111307}-\eqref{constraintS3-20265111307} indicate that the above Hamiltonian analysis is valid only in the case $\bar{A}\ne 0$ and ($\beta_{1}\ne 0$ or $\beta_{4}\ne 0$). Consequently, within the parameter subspace $\bar{A}\ne 0$ and ($\beta_{1}\ne 0$ or $\beta_{4}\ne 0$), the general Einstein-vector theory inevitably violates the scalar sector stability conditions at the linear level. Therefore, this branch of the theory is dynamically unstable under linear scalar perturbations.
}
\par
{We next consider the stability properties in the remaining regions of parameter space. For the case $\bar{A}\ne 0$ and $\beta_{1}=\beta_{4}=0$, the background equation~\eqref{EqA-BG} implies $\mu_{0}=\beta_{2}=0$ since the Hubble parameter $H(t)$ is time dependent. In this case, the action~\eqref{action-all} reduces to that of the Einstein-Maxwell theory supplemented by a Gauss-Bonnet term. The Hamiltonian analysis shows that, apart from the single dynamical scalar perturbation associated with matter, the scalar sector contains only a gauge degree of freedom and no additional propagating scalar modes. Consequently, the only remaining stability requirement is the matter-sector condition $\bar{p}_{m,n}\ge 0$~\eqref{stabilityS2-20265121600}. For the case $\bar{A}= 0$, when $Q_{\bar{A}}=0$, the same conclusion is obtained as in the parameter subspace $\bar{A}\ne 0$ and $\beta_{1}=\beta_{4}=0$. We therefore focus on the remaining branch $\bar{A}= 0$ and $Q_{\bar{A}}\ne 0$.
}
\par
{For the parameter subspace $\bar{A}= 0$ and $Q_{\bar{A}}\ne 0$, the Hamiltonian analysis shows that all constraints are second class:
\begin{eqnarray}
  && P_{\phi_{h}}\approx 0,\quad P_{\varphi_{h}}\approx 0,\quad P_{\phi_{a}}\approx 0,\quad P_{\varphi_{m}}\approx 0,\quad  P_{\phi_{\ell}}\approx 0,\quad P_{\phi_{m}}-\phi_{\ell}\approx 0,\label{constraintSs1-20265111307}\\
  && -3 a^3H^2\phi_{h} +a H\vec{k}^2\varphi_{h} -4\pi G\bar{\rho}_{m,n}\phi_{m} \approx 0,\quad 16\pi G P_{\varphi_{a}} -a^3Q_{\bar{A}}\phi_{a} \approx 0, \label{constraintSs2-20265111307}\\
  && 4\pi G\bar{\rho}_{m,n}\left(\bar{J}\varphi_{h} +\varphi_{m}\right) +a^3H\phi_{h}\approx 0,\quad -\bar{J}\phi_{\ell} +\bar{\rho}_{m,n}\left(\bar{J}\varphi_{h} +\varphi_{m}\right) \approx 0.\label{constraintSs3-20265111307}
\end{eqnarray}
Eliminating the constrained variables yields an effective Hamiltonian involving only the physical dynamical degrees of freedom, together with the corresponding effective Lagrangian. In the small-scale limit, they take the form
\begin{eqnarray}
  H_{\vec{k}\rightarrow\infty}^{(s2)} &=& \frac{8\pi G}{a^3Q_{\bar{A}}}\left(P_{\varphi_{a}}\right)^2 +\frac{a(Q_{\bar{A}}+4\beta_{2}\dot{H})}{32\pi G}\vec{k}^2(\varphi_{a})^2 +\frac{\bar{J}}{2 a^2 \bar{\rho}_{m,n}}\vec{k}^2\left(P_{\phi_{m}}\right)^2 +\frac{\bar{p}_{m,n}}{2 \bar{J}}(\phi_{m})^2, \\
  \mathcal{L}_{s2,\vec{k}\rightarrow\infty} &=& \frac{a^3Q_{\bar{A}}}{32\pi G}(\dot{\varphi}_{a})^2 -\frac{a(Q_{\bar{A}}+4\beta_{2}\dot{H})}{32\pi G}\vec{k}^2(\varphi_{a})^2 +\frac{a^2 \bar{\rho}_{m,n}}{2 \bar{J}}\left(\frac{\dot{\phi}_{m}}{|\vec{k}|}\right)^2 -\frac{\bar{p}_{m,n}}{2 \bar{J}}\vec{k}^2\left(\frac{\phi_{m}}{|\vec{k}|}\right)^2.\label{Ls2-2026513923}
\end{eqnarray}
From the Hamiltonian $H_{\vec{k}\rightarrow\infty}^{(s2)}$, stability requires
\begin{eqnarray}
  Q_{\bar{A}} &>& 0,\label{stabilitySs1-20265122155}\\
  \bar{p}_{m,n} &\ge& 0.\label{stabilitySs2-20265122155}
\end{eqnarray}
The second condition corresponds to the stability of matter perturbations. The first condition guarantees the positivity of the kinetic term associated with $\left(P_{\varphi_{a}}\right)^2$. Furthermore, the stability analysis of vector perturbations in the general Einstein-vector theory requires $\beta_{2}\le 0$. Therefore, once condition~\eqref{stabilitySs1-20265122155} is satisfied, the coefficient of the $(\varphi_{a})^2$ term in the Hamiltonian automatically remains positive in the small-scale limit. Hence, no additional scalar-sector instability arises in this branch of the parameter space.
}
\par
{In this subsection, we have analyzed the stability of linear scalar perturbations in the general Einstein-vector theory under the gauge choice $\alpha = E = 0$. Throughout the analysis, no additional conditions were imposed beyond the assumptions $|\beta_{1}|,|\beta_{2}|,|\beta_{4}|\ll 1$, together with the small-scale limit. Stability requires that the Hamiltonian be bounded from below. Our results show that the theory becomes unstable in the parameter region $\bar{A}\ne 0$ and ($\beta_{1}\ne 0$ or $\beta_{4}\ne 0$), since the scalar-sector Hamiltonian cannot satisfy the positivity conditions in this branch. For the parameter subspaces (i) $\bar{A}\ne 0$ and $\beta_{1}=\beta_{4}=0$, and (ii) $\bar{A}= 0$ and $Q_{\bar{A}}=0$, the scalar sector contains only a single dynamical degree of freedom associated with matter perturbations. In these two branches, the stability condition reduces to $\bar{p}_{m,n} \ge 0$. By contrast, in the parameter subspace $\bar{A}= 0$ and $Q_{\bar{A}}\ne 0$, the scalar sector contains not only the matter degree of freedom but also an additional propagating scalar mode originating from the gravitational sector. In this case, the stability conditions become $Q_{\bar{A}} > 0$~\eqref{stabilitySs1-20265122155} and $\bar{p}_{m,n} \ge 0$~\eqref{stabilitySs2-20265122155}. Therefore, without introducing further constraints on the theory parameters, the linear scalar perturbations remain stable only in the three parameter branches: (i) $\bar{A}\ne 0$ and $\beta_{1}=\beta_{4}=0$, (ii) $\bar{A}= 0$ and $Q_{\bar{A}}=0$, and (iii) $\bar{A}= 0$ and $Q_{\bar{A}}\ne 0$. Among them, only branch (iii) supports a nontrivial propagating gravitational scalar degree of freedom, whereas branches~(i) and~(ii) effectively reduce to matter-only scalar dynamics.}

\subsection{Gravitational waves in the small-scale limit}
{In the previous subsection, we analyzed the stability conditions of scalar perturbations in the general Einstein-vector theory. Since GWs detected by current observatories correspond to high-wavenumber modes compared to the cosmic scale, we now focus on the propagation properties of GWs in the small-scale limit ($|\vec{k}|\rightarrow \infty$), adopting the gauge choice $\alpha=0$ and $E=0$.
}
\par
{We first consider the case $\bar{A}= 0$ and $Q_{\bar{A}}\ne 0$. According to the effective Lagrangian~\eqref{Ls2-2026513923} in the small-scale limit, it is straightforward to see that there are two dynamical scalar perturbations, $\varphi_{a}$ and $\phi_{m}$, in the general Einstein-vector theory. The remaining scalar perturbations are nondynamical and can be expressed through constraint equations~\eqref{constraintSs1-20265111307}-\eqref{constraintSs3-20265111307}, the perturbation $\phi_{a}$ depends on $\varphi_{a}$, while $\phi_{h}$, $\varphi_{h}$, $\varphi_{m}$, and $\phi_{\ell}$ depend on $\phi_{m}$. According to the geodesic deviation equation, GWs are directly determined by perturbations of the metric, which correspond to $\phi_{h}$ and $\varphi_{h}$ among the scalar perturbations. Therefore, only the dynamical scalar perturbation $\phi_{m}$ can contribute indirectly to GW observables, whereas $\varphi_{a}$ does not enter the metric perturbations relevant for GW propagation. However, in vacuum ($\phi_{m}=\varphi_{m}=\phi_{\ell}=0$), the constraint equations~\eqref{constraintSs2-20265111307} and \eqref{constraintSs3-20265111307} imply that $\phi_{h}=\varphi_{h}=0$. This shows that the metric scalar perturbations cannot propagate in vacuum. Consequently, no propagating scalar GW mode exists in this case.
}
\par
{Next, we consider the cases where $\bar{A}\ne 0$ and $\beta_{1}=\beta_{4}=0$, and where $\bar{A}= 0$ and $Q_{\bar{A}}=0$. In both situations, the theory contains only one dynamical scalar perturbation, namely $\phi_{m}$. As in the previous case, the vacuum constraints again enforce $\phi_{h}=\varphi_{h}=0$, indicating the absence of propagating scalar GW modes.
}
\par
{Finally, we turn to the case $\bar{A}\ne 0$ and ($\beta_{1}\ne 0$ or $\beta_{4}\ne 0$). In the most general situation, without imposing additional assumptions beyond $|\beta_{1}|,|\beta_{2}|,|\beta_{4}|\ll 1$ and the small-scale limit, the theory generically fails to satisfy the full stability conditions. However, if one restricts attention to the stability of the plane-wave sector, the resulting conditions are significantly relaxed and may be satisfied in certain parameter regions. From the Hamiltonian~\eqref{Hs1-20265131119}, one observes that the canonical coordinate $\phi_{1}$ does not appear explicitly in the Hamiltonian. Therefore, its conjugate momentum is a conserved quantity, $\dot{P}_{\phi_{1}}=\{ P_{\phi_{1}},H_{\vec{k}\rightarrow\infty}^{(s1)} \}=0$. Considering plane-wave solutions, we obtain $P_{\phi_{1}}=0$. Substituting this result back into the Hamiltonian~\eqref{Hs1-20265131119}, the stability condition $H_{\vec{k}\rightarrow\infty}^{(s1)}\ge 0$, to leading order in $\beta_{\bullet}$, reduces to
\begin{eqnarray}
  \frac{16\beta_{4}\dot{H}}{3\beta_{1}+12\beta_{4}H^2} &\ge& -1,\label{stab-20265132059}\\
  \bar{p}_{m,n} &\ge& 0.
\end{eqnarray}
Therefore, if one is concerned only with the stability of the plane GW system, the Hamiltonian remains positive definite provided that the above conditions are satisfied. This reduction indicates that, although the full system may generically suffer from instabilities, the plane-wave sector can still admit a consistent and stable regime for suitable parameter choices.
}
\par
{Next, we investigate the properties of scalar GWs in the case $\bar{A}\ne 0$ and ($\beta_{1}\ne 0$ or $\beta_{4}\ne 0$). Imposing the plane-wave condition $P_{\phi_{1}}=0$, the corresponding effective Lagrangian in the small-scale limit becomes
\begin{eqnarray}
  \mathcal{L}_{s1,\vec{k}\rightarrow\infty}&=& \frac{3a^3H^2q_{t}}{16\pi G}\left(\big(\dot{\phi}_{2}\big)^2 -c_{s}^2\frac{\vec{k}^2}{a^2}\phi_{2}^2\right) +\frac{a^2\bar{\rho}_{m,n}}{2\bar{J}\vec{k}^2}\left(\big(\dot{\phi}_{m}\big)^2 -\frac{\bar{p}_{m,n}}{\bar{\rho}_{m,n}}\frac{\vec{k}^2}{a^2}\phi_{m}^2\right),\label{eLs1-20265132026}
\end{eqnarray}
where
\begin{eqnarray}
  c_{s}^2=\frac{F_{7}(t)}{3 a H^2q_{t}} -\frac{\big( q_{t} -4\bar{A}^2\big((1-2\beta_{2})(\beta_{1}+4\beta_{4}H^2)+4\beta_{4}\dot{H}\big) \big)^2}{12\bar{A}^2(\beta_{1}+4\beta_{4}H^2)(1-2\beta_{1}-8\beta_{4}H^2)q_{t}}.
\end{eqnarray}
The above Lagrangian shows that, in the small-scale limit, the scalar sector effectively decomposes into two propagating modes: the matter perturbation $\phi_{m}$ and the additional scalar mode $\phi_{2}$. The positivity of the square of the propagation velocity $c_{s}^2$ is therefore required to avoid gradient instabilities in the scalar GW sector, i.e., the condition~\eqref{stab-20265132059}. Furthermore, from the constraint equations~\eqref{constraintS1-20265111307}-\eqref{phi2-20265111931}, it is straightforward to show that the metric perturbations $\phi_{h}$ and $\varphi_{h}$ depend on both $\phi_{2}$ and $\phi_{m}$. In vacuum, however, they depend only on $\phi_{2}$, since the matter perturbations vanish. Consequently, unlike the previous cases, the theory now admits one propagating scalar degree of freedom associated with GWs, namely $\phi_{2}$.}
\par
Varying the Lagrangian~\eqref{eLs1-20265132026} with respect to $\phi_{2}$ and $\phi_{m}$ yields their respective dispersion relations in the small-scale limit:
\begin{eqnarray}
   w_{\phi_{2}}^2 -c_{s}^2\bar{g}^{ij}k_{i}k_{j}&=&0,\\
   w_{\phi_m}^2 -\frac{\bar{p}_{m,n}}{\bar{\rho}_{m,n}}\bar{g}^{ij}k_{i}k_{j}&=&0.
\end{eqnarray}
Here, $\bar{p}_{m,n}/\bar{\rho}_{m,n}$ represents the squared matter sound speed. The squared propagation speed $c_s^2$ of scalar GWs can be expressed as
\begin{eqnarray}
  c_{s}^2= 1 -\frac{\beta_{2}}{3(\beta_{1}+4\beta_{4}H^2)q_{t}}(\dots) +\frac{\beta_{4}}{3(\beta_{1}+4\beta_{4}H^2)q_{t}} ( \dots ).\label{cs2-expand-2138}
\end{eqnarray}
Obviously, whether the propagation speed of scalar GWs deviates from the speed of light depends on whether the parameters $\beta_{2}$ and $\beta_{4}$ vanish. According to Eq.~\eqref{tensorialGW-constraint-1528}, if the speed of scalar GWs coincides with the speed of light, namely $\beta_{2}=\beta_{4}=0$, then the propagation speed of tensor GWs is exactly equal to the speed of light.
\par
~\

{In this section, we have analyzed the dynamics of scalar perturbations in the general Einstein-vector theory under the gauge condition $\alpha=E=0$, together with the small-scale limit. We first investigated the stability of linear scalar perturbations, which requires the Hamiltonian to be positive definite. For the parameter subspace $\bar{A}\ne 0$ and ($\beta_{1}\ne 0$ or $\beta_{4}\ne 0$), in addition to one dynamical degree of freedom arising from matter perturbations, the theory contains two additional dynamical degrees of freedom in the scalar gravitational sector. We found that, without imposing further restrictions beyond the assumptions $|\beta_{1}|,|\beta_{2}|,|\beta_{4}|\ll 1$, the full stability conditions generally cannot be satisfied in this case. However, if one focuses only on the stability of the plane-wave sector, the stability conditions become significantly less restrictive and can be satisfied under the condition given in Eq.~\eqref{stab-20265132059}. For the parameter subspace $\bar{A}\ne 0$ and $\beta_{1}=\beta_{4}=0$, as well as for the parameter subspace $\bar{A}= 0$ and $Q_{\bar{A}}=0$, the theory possesses only one dynamical degree of freedom arising from matter perturbations. In these cases, the stability condition reduces simply to $\bar{p}_{m,n} \ge 0$. For the parameter subspace $\bar{A}= 0$ and $Q_{\bar{A}}\ne 0$, besides one matter dynamical degree of freedom, the theory contains an extra dynamical scalar degree of freedom originating from the gravitational sector. In this case, stability requires both $Q_{\bar{A}} > 0$~\eqref{stabilitySs1-20265122155} and $\bar{p}_{m,n} \ge 0$~\eqref{stabilitySs2-20265122155}. We then investigated the propagation properties of scalar GWs and found that propagating scalar GW modes arise only in the parameter subspace $\bar{A}\ne 0$ and ($\beta_{1}\ne 0$ or $\beta_{4}\ne 0$). In the small-scale limit and for plane-wave solutions, the scalar GW sector contains a single independent propagating mode with, in general, a nonluminal propagation speed. This propagation speed reduces to the speed of light in the special case $\beta_{2}=0$ and $\beta_{4}=0$. The main conclusions of this analysis are summarized in Table~\ref{scalar-conclusion-181017}.}
\begin{table}[h]
\centering
\begin{tabular}{|c|c|c|c|c|}
  \hline
  \textbf{Perturbations} & \textbf{Case} & \textbf{d.o.f.} & \textbf{Stability} & \textbf{Number of GW modes}
  \\ \hline
  \multirow{3}*{Scalar} 
      & $\bar{A}\ne 0$ and ($\beta_{1}\ne 0$ or $\beta_{4}\ne 0$) & 3 & unstable & 1 
      \\ \cline{2-5}
      & $\bar{A}=0$ and $Q_{\bar{A}}\ne 0$ & 2 & $Q_{\bar{A}}>0$,$\bar{p}_{m,n} \ge 0$ & 0 
      \\ \cline{2-5}
      & \makecell{$\bar{A}\ne 0$ and $\beta_{1}=\beta_{4}=0$, \\ or $\bar{A}=0$ and $Q_{\bar{A}}=0$} & 1 & $\bar{p}_{m,n} \ge 0$ & 0 
  \\ \hline
\end{tabular}
\caption{{Dynamics of scalar perturbations in the general Einstein-vector theory. The conclusions are derived in the small-scale limit under the assumption $|\beta_{1}|,|\beta_{2}|,|\beta_{4}|\ll 1$. The column labeled ``d.o.f." denotes the number of dynamical degrees of freedom. In general, the full stability conditions cannot be satisfied in the case $\bar{A}\ne 0$ and ($\beta_{1}\ne 0$ or $\beta_{4}\ne 0$). However, if the plane-wave solution condition is additionally imposed, the stability condition becomes relaxed and can be satisfied.}}
\label{scalar-conclusion-181017}
\end{table}

\section{Conclusion}\label{conclusion-sec.7-1132252}
The general Einstein-vector theory~\cite{Geng:2015kvs} is an extension of Einstein-Maxwell theory that introduces a mass term and additional couplings between the vector field $A_{\mu}$ and curvature tensors. As a result, the extended theory no longer possesses the $U(1)$ gauge symmetry associated with the vector field. However, an approximate and emergent gauge symmetry can arise at the linear perturbative level on backgrounds in which $\bar{A}$ vanishes. This emergent symmetry has negligible experimental or observational consequences in the solar system. By contrast, on large scales or in cosmological settings, it can give rise to a variety of nontrivial effects that may be testable by future observations. In the context of cosmic evolution, the vector field can play a role of the inflaton, and there exist solutions in which the inflaton vanishes at late times~\cite{Geng:2015kvs}. Moreover, the general Einstein-vector theory is an intriguing candidate for explaining dark energy and dark matter. The distinctive features of this theory also lead to a rich spectrum of GW phenomena. Consequently, studying this theory provides an important theoretical framework for future cosmological observations and GW detection.
\par
{In this paper, we investigated the stability and GW properties of the four-dimensional general Einstein-vector theory on a cosmological background. Under the assumption $|\beta_{1}|,|\beta_{2}|,|\beta_{4}|\ll 1$, we analyzed the stability using the Hamiltonian formalism at the linear perturbative level, where stability requires the Hamiltonian to be positive definite. For the tensor perturbations, the stability conditions can be readily satisfied. For the vector perturbations, stability requires $\beta_{2}\le 0$ when $\bar{A}\ne 0$. The scalar sector exhibits a significantly richer structure. For the parameter subspace $\bar{A}\ne 0$ and ($\beta_{1}\ne 0$ or $\beta_{4}\ne 0$), the full stability conditions generally cannot be satisfied unless the plane-wave condition is additionally imposed. By contrast, for the parameter subspace $\bar{A}\ne 0$ and $\beta_{1}=\beta_{4}=0$, as well as for the parameter subspace $\bar{A}= 0$ and $Q_{\bar{A}}=0$, the theory contains only one dynamical scalar degree of freedom associated with matter perturbations, with the stability condition $\bar{p}_{m,n} \ge 0$. For the parameter subspace $\bar{A}= 0$ and $Q_{\bar{A}}\ne 0$, besides one matter dynamical degree of freedom, the theory possesses an extra dynamical scalar degree of freedom originating from the gravitational sector. In this case, stability further requires $Q_{\bar{A}} > 0$. The main results are summarized in Tables~\ref{tensor-conclusion-181017},~\ref{vector-conclusion-181017}, and~\ref{scalar-conclusion-181017}. Note that the stability conditions for the scalar perturbations listed in Table~\ref{scalar-conclusion-181017} are necessary but not sufficient, as they are obtained in the small-scale limit. }
\par
Furthermore, in the small-scale limit ($|\vec{k}|\rightarrow\infty$), we investigated the GW properties of the general Einstein-vector theory. For tensor GWs, there exist two propagating modes. Based on the constraint from the GW event GW170817 and its electromagnetic counterpart GRB170817A, we can essentially assume that tensor GWs propagate at the speed of light. This requirement restricts the parameter space to the following three cases: i) $\bar{A}=0$, ii) $\bar{A}=$const. with $\beta_{2}=0$, and iii) $\beta_{2}=\beta_{4}=0$. For vector GWs, there are two propagating modes with superluminal speeds when $\beta_{2}\ne 0$ and $\bar{A}\ne 0$, whereas no vector GWs propagate when $\beta_{2}= 0$ or $\bar{A}= 0$. For scalar GWs, in the case $\bar{A}\ne 0$ and ($\beta_{1}\ne 0$ or $\beta_{4}\ne 0$), there exists a single propagating mode, otherwise, they are absent. The propagation speed of scalar GWs coincides with the speed of light only when $\beta_{2}=0$ and $\beta_{4}=0$. These results were summarized in Tables~\ref{tensor-conclusion-181017},~\ref{vector-conclusion-181017}, and~\ref{scalar-conclusion-181017}.  {In particular, without imposing the plane wave condition, no scalar GW modes can exist within the stable parameter space. Furthermore, we found that when tensor GWs propagate strictly at the speed of light, the general Einstein-vector theory forbids the existence of vector GWs. These distinctive features provide a potentially powerful observational test of the theory in future GW experiments. These results demonstrate that the stability and propagation properties of perturbative modes depend sensitively on both the background vector configuration and the coupling parameters of the theory. Specifically, the scalar sector imposes the strongest restrictions on the viable parameter space and plays a crucial role in determining whether additional GW polarizations can propagate consistently.}
\par
Many researches exist on related aspects. In Ref.~\cite{Lai:2024fza}, the polarization modes of GWs in the general Einstein-vector theory in a Minkowski background were examined, omitting terms involving $\beta_{4}$. Under the same assumption, namely $\beta_{4}=0$, we found that our results are broadly consistent with those reported in the Ref.~\cite{Lai:2024fza}. However, the present analysis leads to more restrictive conclusions. Given that the current universe is undergoing accelerated expansion, the case $\bar{A}\ne 0$ with $\beta_{1}=\beta_{4}=0$ does not allow for the existence of scalar GWs. By contrast, Ref.~\cite{Lai:2024fza} considered a Minkowski background, under which scalar GWs may still propagate. Moreover, by incorporating stability requirements, our analysis imposes additional constraints on the propagation speeds of GWs. Regarding stability, owing to the structural similarity between the general Einstein-vector theory and Bumblebee theory, their stability conditions are expected to be closely related. The stability of Bumblebee theory has been investigated in Refs.~\cite{vandeBruck:2025aaa} and~\cite{Lai:2025nyo}. A direct comparison of the corresponding actions shows that the cosmological constant term $-2\Lambda_{0}$, together with the vector mass term $-\mu_{0}^2A^2/2$ in the general Einstein-vector theory, corresponds to a specific choice of the potential term $V(B_{\mu}B^{\mu}\pm b^2)$ in Bumblebee theory. Consequently, for $\beta_{4}=0$, the two theories are expected to yield similar results in their stability analyses. When $\beta_{4}\ne 0$, however, they exhibit fundamentally different behaviors with respect to the number of dynamical degrees of freedom, the propagation speeds of perturbations, and their stability properties. Notably, these differences manifest primarily in the scalar sector, as summarized in Table~\ref{scalar-conclusion-181017}.
\par
The general Einstein-vector theory has rich implications for cosmological evolution, dark matter, dark energy, and GWs. Our work provides an alternative theoretical perspective on understanding the current cosmic dynamics and GWs properties within the broader class of vector-tensor theories. With the continuous detection of ground-based GW detectors, such as LIGO, Virgo, KAGRA, as well as PTAs and FAST~\cite{LIGOScientific:2007fwp,VIRGO:2012dcp,KAGRA:2018plz,NANOGrav:2023gor,EPTA:2023fyk,Reardon:2023gzh,Xu:2023wog,Qian_2020}, together with the rapid progress of space-based missions including LISA, Taiji, and TianQin~\cite{Luo:2021qji, TianQin:2015yph, LISA:2017pwj}, the distinctive GW signatures predicted by this theory are expected to be tested in the near future. These signatures include the polarization modes, propagation speeds, and the correlations between the tensor, vector, and scalar modes. Furthermore, the symmetry and the dynamics of this theory may also be probed observationally by forthcoming cosmological and GW experiments.

\section*{Acknowledgments}
We would like to thank Shan-Ping Wu for useful discussions. 
This work is supported in part by the National Key Research and Development Program of China (Grant No. 2021YFC2203003), 
the National Natural Science Foundation of China (Grants No. 12475056, No. 123B2074, and No. 12247101), Gansu Province's Top Leading Talent Support Plan, the Fundamental Research Funds for the Central Universities (Grant No. lzujbky-2025-jdzx07), the Natural Science Foundation of Gansu Province (No. 22JR5RA389 and No. 25JRRA799), the 111 Project (Grant No. B20063), {and the Department of Education of Gansu Province: Outstanding Graduate ``Innovation Star” Project (Grant No. 2026CXZX-044)}.\par

\appendix

\section{The general Einstein-vector theory}\label{GEV-theory-1171300}
The general Einstein-vector theory is a vector-tensor theory formulated in arbitrary spacetime dimensions $D$, originally constructed by Lu and Geng in 2015~\cite{Geng:2015kvs}. In addition to the spacetime metric $g_{\mu\nu}$, the theory contains a vector field $A^{\mu}$ that couples bilinearly to curvature polynomials of arbitrary order. These couplings are arranged such that only the Riemann tensor, and not its derivatives, appears in the resulting equations of motion. Moreover, the equation of motion for the vector field is linear in $A^{\mu}$ and involves at most second derivatives. Consequently, the general Einstein-vector theory belongs to the class of second-order derivative gravity theories.
\par
The complete Lagrangian for the general Einstein-vector theory is given by~\cite{Geng:2015kvs}
\begin{equation}
    \mathcal{L}=\sqrt{-g}\left( -\frac{1}{4}F^2+\sum_{k=0}\left( \alpha^{(k)}E^{(k)}+\beta^{(k)}\widetilde{G}^{(k)}
    +\gamma^{(k)}G^{(k)} \right) \right),\label{lagrangian_GEV}
\end{equation}
where $F_{\mu\nu}=\nabla_{\mu}A_{\nu}-\nabla_{\nu}A_{\mu}$ denotes the field-strength tensor associated with the vector potential $A^{\mu}$, and $F^2=F_{\mu\nu}F^{\mu\nu}$. Here, $\alpha^{(k)}$, $\beta^{(k)}$, $\gamma^{(k)}$ are sets of constant parameters, while $E^{(k)}$, $\widetilde{G}^{(k)}$, and $G^{(k)}$ are defined as
\begin{eqnarray}
    && E^{(k)}=\frac{1}{2^k}\delta^{\beta_1 \cdots \beta_{2k}}_{\alpha_1 \cdots \alpha_{2k}}
    R^{\alpha_1\alpha_2}_{\quad\;\;\;\beta_1\beta_2} \cdots R^{\alpha_{2k-1}\alpha_{2k}}_{\qquad\quad\;\;\beta_{2k-1}\beta_{2k}},\\
    && \widetilde{G}^{(k)}=E^{(k)}A^2,\\
    && G^{(k)}=E^{(k)}_{\mu\nu}A^{\mu}A^{\nu}.
\end{eqnarray}
Here, $E^{(k)\nu}_{\mu}=-\frac{1}{2^{k+1}}\delta^{\beta_1 \cdots \beta_{2k}\nu}_{\alpha_1 \cdots \alpha_{2k}\mu}
R^{\alpha_1\alpha_2}_{\quad\;\;\;\beta_1\beta_2} \cdots R^{\alpha_{2k-1}\alpha_{2k}}_{\qquad\quad\;\;\beta_{2k-1}\beta_{2k}}$, $R^{\mu\nu}_{\quad\alpha\beta}$ is Riemann tensor, $\delta^{\beta_1 \cdots \beta_{s}}_{\alpha_1 \cdots \alpha_{s}}=s!\delta^{\beta_1}_{[\alpha_1} \cdots \delta^{\beta_s}_{\alpha_s]}$, and 
$A^2=A_{\mu}A^{\mu}$. In the theory described by Eq.~\eqref{lagrangian_GEV}, it is straightforward to see that setting $A^{\mu}=0$ reduces the theory to pure Lovelock gravity.
\par
In this paper, we focus on the four-dimensional case ($D=4$). In this dimension, all terms with $k>2$ in the Lagrangian~\eqref{lagrangian_GEV} vanish, so the action reduces to
\begin{eqnarray}
    S_{g}&=&\frac{1}{16\pi G}\int d^{4}x\sqrt{-g} \Big[ \alpha^{(1)}R+\alpha^{(0)}-\frac{1}{4}F^2+ \left(\beta^{(0)}-\frac{\gamma^{(0)}}{2}\right)A^2 +\beta^{(1)}R A^2+\gamma^{(1)}G_{\mu\nu}A^{\mu}A^{\nu}\nonumber\\
    && +\alpha^{(2)}E^{(2)}+\beta^{(2)}E^{(2)}A^2 \Big], \label{action4-initial}
\end{eqnarray}
where $G_{\mu\nu}=R_{\mu\nu}-\frac{1}{2}g_{\mu\nu}R$ is the Einstein tensor, and $E^{(2)}=R^{2}-4R^{\mu\nu}R_{\mu\nu}+R^{\mu\nu\alpha\rho}R_{\mu\nu\alpha\rho}$ is the Gauss-Bonnet term.
\par
By comparing the action in Eq.~\eqref{action4-initial} with that of Einstein's GR, we can rewrite it as
\begin{eqnarray}
    S_{g}&=&\frac{1}{16\pi G}\int d^{4}x\sqrt{-g} \left[ R-2\Lambda_{0}-\frac{1}{4}F^2-\frac{\mu_0^2}{2}A^2 +\beta_1R A^2+\beta_2G_{\mu\nu}A^{\mu}A^{\nu}+\beta_{3}E^{(2)}+\beta_{4}E^{(2)}A^{2} \right]. \label{g-action-4d}
\end{eqnarray}
Here, $\Lambda_{0}$ is the cosmological constant, $\mu_{0}$ is the vector field mass, and $\beta_1,\dots,\beta_4$ are coupling constants. Since the term $\beta_{3}E^{(2)}$ corresponds to the pure Gauss-Bonnet term, it does not contribute to the equations of motion.

\section{The Schutz-Sorkin action} \label{appendix-Schutz-Sorkin-10162254}
In its rest frame, a perfect fluid is uniquely characterized by its energy density and pressure. For a perfect fluid that does not couple explicitly to the curvature, it is natural to choose either the energy density $\rho$ ($\mathcal{L}_m=-\rho$)~\cite{Brown:1992kc, hawking2023large} or the pressure $p$ ($\mathcal{L}_m=p$)~\cite{Brown:1992kc, Schutz:1970my} as the matter Lagrangian density. Another admissible choice is $\mathcal{L}_m=-na$~\cite{Brown:1992kc, Bertolami:2008ab}, where $n$ is the particle number density and $a$ is the physical free energy per particle, defined by $a=\rho/n-Ts$, with $T$ denoting the temperature and $s$ the entropy per particle. These three Lagrangian densities are equivalent within the framework of GR~\cite{Brown:1992kc}. When matter couples nonminimally to the Ricci scalar, several studies have investigated such couplings~\cite{Bertolami:2008ab, Faraoni:2009rk}. For further discussions of perfect-fluid Lagrangians, see Refs.~\cite{deBoer:2017ing, Ovalle:2017fgl, Buchert:2001sa}.
\par
In this paper, we focus on a minimally coupled perfect fluid described by the Schutz-Sorkin action~\cite{Schutz:1977df, DeFelice:2009bx, Bertolami:2008ab, DeFelice:2016yws, Kase:2018nwt, Lai:2025nyo}
\begin{eqnarray}
  S_m=-\int d^4x \left[ \sqrt{-g}\rho_m(n)+J^{\mu}(\partial_{\mu}\ell+\mathcal{A}_1\partial_{\mu}\mathcal{B}_1+\mathcal{A}_2\partial_{\mu}\mathcal{B}_2) \right]. \label{perfect-fluid-action}
\end{eqnarray}
Here, $\rho_{m}$ is the energy density, $n$ the particle number density, $J^{\mu}$ a vector density, and $\ell$ a scalar. The quantities $\mathcal{A}_1$, $\mathcal{A}_2$, $\mathcal{B}_1$, and $\mathcal{B}_2$ arise from the intrinsic vector perturbations of the matter (see Refs.~\cite{DeFelice:2009bx, DeFelice:2016yws}).
\par 
Note that the matter action $S_m$ is a functional of $g_{\mu\nu}$, $J^{\mu}$, $\ell$, $\mathcal{A}_1$, $\mathcal{A}_2$, $\mathcal{B}_1$, and $\mathcal{B}_2$, i.e., 
\begin{eqnarray}
  S_m=S_m[g_{\mu\nu},J^{\mu},\ell,\mathcal{A}_1,\mathcal{A}_2,\mathcal{B}_1,\mathcal{B}_2].
\end{eqnarray}
The scalar field $\ell$ acts as a Lagrange multiplier enforcing the constraint $\partial_{\mu}J^{\mu}=0$, which expresses particle-number conservation. The vector density $J^{\mu}$, representing the particle-number flux, is defined in terms of the number density $n$ and the four-velocity $U^\mu$ as
\begin{eqnarray}
    J^{\mu}=\sqrt{-g}nU^{\mu}.\label{vectorDensity-all}
\end{eqnarray}
The four-velocity satisfies the normalization $U^{\mu}U_{\mu}=-1$. The particle number density is then given by $n=|J|/\sqrt{-g}$. Consequently, the energy density is a function of this quantity: $\rho_{m}=\rho_{m}(|J|/\sqrt{-g})$. 
\par
Varying the action~\eqref{perfect-fluid-action} with respect to the metric $g_{\mu\nu}$ yields the perfect-fluid energy-momentum tensor
\begin{eqnarray}
    T^{\mu\nu}=\rho_m U^{\mu}U^{\nu}+\left( n\frac{\partial\rho_m}{\partial n}-\rho_m \right)\left(g^{\mu\nu}+U^{\mu}U^{\nu}\right).
\end{eqnarray}
Here we adopt the standard definition of the matter energy-momentum tensor $T_{\mu\nu}=-\frac{2}{\sqrt{-g}}\frac{\delta (\sqrt{-g}\mathcal{L}_m)}{\delta (g^{\mu\nu})}$. We now consider the energy-momentum tensor of a perfect fluid, $T^{\mu\nu}=(\rho_m +p_m)U^{\mu}U^{\nu}+p_mg^{\mu\nu}$. By comparing these two expressions, the pressure can be identified as
\begin{eqnarray}
    p_m=n\frac{\partial\rho_m}{\partial n}-\rho_m. \label{pm-2127}
\end{eqnarray}
\par
Varying the action~\eqref{perfect-fluid-action} with respect to the vector density $J^{\mu}$, and noting that the gravitational action $S_g$ is independent of $J^{\mu}$, yields
\begin{eqnarray}
  U_{\mu}\equiv \frac{J_{\mu}}{|J|}=\frac{1}{\rho_{m,n}}\left(\partial_{\mu}\ell+\mathcal{A}_1\partial_{\mu}\mathcal{B}_1+\mathcal{A}_2\partial_{\mu}\mathcal{B}_2\right),
\end{eqnarray}
where $\rho_{m,n}=\partial\rho_m/\partial n$. One can show that the spatial components $U_i$ of $U_{\mu}$ can be decomposed into a scalar part and a divergence-free vector part. This decomposition remains valid even when $\rho_{m,n}$ is constant, in agreement with Refs.~\cite{DeFelice:2009bx, Schutz:1970my}. In a cosmological background, the divergence-free vector component of $U_i$ is sourced by the scalar variables $\mathcal{A}_1$, $\mathcal{A}_2$, $\mathcal{B}_1$, and $\mathcal{B}_2$.

\section{The specific forms of some quantities}\label{quantities-2314}
This appendix details the specific forms of the complex quantities referenced throughout the paper.
\par
Explicit expressions for key quantities in the scalar perturbation action~\eqref{action-second-1030} are:
\begin{eqnarray}
    Q_{\bar{A}}&=&\mu_{0}^2 -12\beta_{1}\big(2H^2+\dot{H}\big) +6\beta_{2}H^2 -48\beta_{4}H^2\big(H^2+\dot{H}\big),\\
    Q_{1}&=& \frac{3 a^3}{16\pi G} \Big( -2H^2 -2\beta_{2}\bar{A} \big( (4H^2+3\dot{H})\bar{A}-4H\dot{\bar{A}} \big) +5\beta_{2}H^2\bar{A}^2 +8\beta_{4}H^2\bar{A}\big( 6H\dot{\bar{A}}-5(H^2+\dot{H})\bar{A} \big) \Big),\\
    Q_{2}&=& \frac{a}{4\pi G}\bar{A}(\beta_{1}+4\beta_{4}H^2),\\
    Q_{4}&=& \frac{a}{4\pi G} \Big( -H +\beta_{1}\bar{A}\big(\dot{\bar{A}}-\bar{A}H\big) +\frac{3}{2}\beta_{2}H\bar{A}^2 +4\beta_{4}\bar{A}H \big(3H\dot{\bar{A}}-2(H^2+\dot{H})\bar{A}\big) \Big),\\
    Q_{5}&=& \frac{3 a^3}{4\pi G}\Big( (\beta_{1}+4\beta_{4}H^2)\big( H\dot{\bar{A}}-(H^2+\dot{H})\bar{A} \big) +\beta_{2}H^2\bar{A} \Big),\\
    Q_{6}&=& \frac{a}{4\pi G} \Big( (\beta_{1}+4\beta_{4}H^2)\big( \dot{\bar{A}}-H\bar{A} \big) +\beta_{2}H\bar{A} \Big),\\
    Q_{7}&=& -\frac{a}{4\pi G}\Big( H -3\beta_{1}\bar{A}\dot{\bar{A}} -\frac{3}{2}\beta_{2}H\bar{A}^2 +4\beta_{4}H^2\bar{A}\big(H\bar{A}-5\dot{\bar{A}}\big) \Big),\\
    Q_{8}&=& \frac{3 a^3}{8\pi G} \Bigg[ -3H^2-2\dot{H} +\beta_{1}\Big( 8H\bar{A}\dot{\bar{A}}+\dot{H}\bar{A}^2+4\dot{\bar{A}}^2+4\bar{A}\ddot{\bar{A}} \Big) +\frac{1}{2}\beta_{2}\Big( 9H^2\bar{A}^2+8H\bar{A}\dot{\bar{A}}+4\dot{H}\bar{A}^2 \Big) \nonumber\\
      && +12\beta_{4}\Bigg(  -H^4\bar{A}^2+4H^3\bar{A}\dot{\bar{A}}+2H^2\Big(\dot{\bar{A}}^2+\bar{A}\ddot{\bar{A}}\Big)-H^2\dot{H}\bar{A}^2+4H\dot{H}\bar{A}\dot{\bar{A}} \Bigg) \Bigg],\\
    Q_{9}&=& \frac{3 a^3}{8\pi G}\Big( -H +3\beta_{1}\bar{A}\dot{\bar{A}} +\frac{3}{2}\beta_{2}H\bar{A}^2 +4\beta_{4}H^2\bar{A}\big( 5\dot{\bar{A}}-H\bar{A} \big) \Big),\\
    Q_{10}&=& \frac{a}{8\pi G}\Big( H -\frac{1}{2}\bar{A}(2\beta_{1}+\beta{2})\big( 2\dot{\bar{A}}+H\bar{A} \big) -\beta_{4}\big( 8H(\dot{\bar{A}}^2+\bar{A}\ddot{\bar{A}}) +8(H^2+\dot{H})\bar{A}\dot{\bar{A}} \big) \Big).
\end{eqnarray}
\par
{The explicit form of the canonical Hamiltonian $H_{C}^{(s)}$~\eqref{HCs-20265111041} for scalar perturbations is
\begin{eqnarray}
  H_{C}^{(s)} &=& \frac{8 \pi G}{a \vec{k}^2} (P_{\varphi_a})^2 + P_{\varphi_a} \phi_a - \left( Q_1 + Q_2 \bar{A}\vec{k}^2  \right) (\phi_h)^2- \frac{\bar{J} \bar{\rho}_{m,n}}{2 a^2} \vec{k}^2 (\varphi_h)^2 - \frac{a^3 Q_{\bar{A}}}{32\pi G} (\phi_a)^2  \nonumber\\
  && + \frac{Q_{\bar{A}}+ 4 \beta_2 \dot{H}}{32\pi G}a \vec{k}^2 (\varphi_a)^2 + \frac{\bar{\rho}_{m,nn}}{2 a^3} (\phi_m)^2 - \frac{\bar{\rho}_{m,n}}{2 \bar{J} a^2} \vec{k}^2 (\varphi_m)^2 + \frac{1}{a^2} \vec{k}^2 \phi_{\ell} \varphi_m + \frac{\beta_2 a H \bar{A}}{4 \pi G} \vec{k}^2 \phi_h \varphi_a \nonumber\\
  && + Q_4 \vec{k}^2 \phi_h \varphi_h + \left( Q_5 + Q_2 \left( \vec{k}^2 - 3 a^2 (2 H^2 + \dot{H}) \right) - 3 a^2 H \dot{Q}_2 \right) \phi_h \phi_a - \left( Q_6 - \dot{Q}_2 \right) \vec{k}^2 \phi_a \varphi_h \nonumber\\
  &&+ \bar{\rho}_{m,n} \phi_h \phi_m - \frac{\bar{\rho}_{m,n}}{a^2} \vec{k}^2 \varphi_h \varphi_m.\label{SHCs-20265111043}
\end{eqnarray}
}
{The explicit form of the quantity $K_{1}$ in the second constraint of Eq.~\eqref{constraintS3-20265111307} is
\begin{eqnarray}
  &&K_{1}\left[ P_{\varphi_{a}},\phi_{h},\varphi_{h},\phi_{a},\varphi_{a},\phi_{m},\varphi_{m} \right]\nonumber\\
  &=& -\bar{A}P_{\varphi_a} -\frac{\beta_2 a H \bar{A}}{4 \pi G} \vec{k}^2 \varphi_a -\bar{\rho}_{m,n}\phi_m + 3 H\bar{\rho}_{m,n} \varphi_m \nonumber\\
  && + \frac{a \bar{A}}{4 \pi G} \left( \vec{k}^2 \bar{A} \left( \beta_1 + 4 \beta_4 H^2 \right) + 3 a^2 \left( H \left( \beta_1 + 4 \beta_4 H^2 \right) \left( \bar{A} H + 3 \dot{\bar{A}} \right) - \bar{A} \left( \beta_1 - 4 \beta_4 H^2 \right) \dot{H} \right) \right) \phi_h \nonumber\\
  && + \frac{1}{8 \pi G} \left( 24 \pi G \bar{J} H \bar{\rho}_{m,n} + \vec{k}^2 a \left( \left( 2 -(2\beta_1 + \beta_2) \bar{A}^2 \right) H -2\bar{A} \left( \beta_1 + 12 \beta_4 H^2 \right) \dot{\bar{A}} \right) \right) \varphi_h \nonumber\\
  && -\frac{a\bar{A}}{16 \pi G} \left( 4 \left( \beta_1 + 4 \beta_4 H^2 \right) \vec{k}^2 - a^2 \left( Q_{\bar{A}} + 24 \left( \beta_1 + 4 \beta_4 H^2 \right)\left( H^2 + \dot{H} \right) \right) \right) \phi_a. \label{K1-20265111308}
\end{eqnarray}
}
{The explicit forms of the coefficients $F_{\bullet}(t,\vec{k})$ in Eqs.~\eqref{phi1-20265111931} and~\eqref{phi2-20265111931} are
\begin{eqnarray}
  F_{2}(t,\vec{k})&=&\bar{A} a (\beta_1 + 4 \beta_4 H^2)^2 \left[ \left(2 + (4\beta_1 - \beta_2 + 24\beta_4 H^2) \bar{A}^2\right) H^2 + 4\beta_1 \bar{A} \dot{\bar{A}} H + 2 \bar{A}^2 (\beta_1 + 12\beta_4 H^2) \dot{H} \right] \vec{k}^6 \nonumber\\
  && -12 \bar{A} (\beta_1 + 4\beta_4 H^2) \Big[ 2 G \bar{J} \pi H^2 (\beta_1 + 4\beta_4 H^2) \bar{\rho}_{m,n} + a^3 H^3 (\beta_1 + 4\beta_4 H^2) \big( 2 H + \bar{A} (5\beta_1 - 2\beta_2 \nonumber\\
  && + 24\beta_4 H^2) (2 \bar{A} H + \dot{\bar{A}}) + (5\beta_1 - \beta_2) \bar{A} \dot{\bar{A}} \big) + 2 a^3 H \big( \beta_1 (1 + (4\beta_1 - \beta_2) \bar{A}^2) H + 4\beta_4 (1 + 2(8\beta_1 \nonumber\\
  &&  - \beta_2) \bar{A}^2) H^3 + 192 \beta_4^2 \bar{A}^2 H^5 + 2 \bar{A} (\beta_1 + 4\beta_4 H^2)(\beta_1 + 6\beta_4 H^2) \dot{\bar{A}} \big) \dot{H} + 2 a^3 \bar{A}^2 (\beta_1 + 8\beta_4 H^2)(\beta_1\nonumber\\
  && + 12\beta_4 H^2) \dot{H}^2 \Big] \vec{k}^4 + 36 \bar{A} a^2 (\beta_1 + 4\beta_4 H^2) \Big[ a^3 (H^2 + \dot{H}) \Big( H^3 (\beta_1 + 4\beta_4 H^2) \big( H (2 + \bar{A}^2 (16\beta_1 \nonumber\\
  && - 7\beta_2 + 72\beta_4 H^2)) + 2 \bar{A} (8\beta_1 - 3\beta_2 + 24\beta_4 H^2) \dot{\bar{A}} \big) + H \Big( \beta_1 (2 + (10\beta_1 - 3\beta_2) \bar{A}^2) H \nonumber\\
  && + 4\beta_4 (2 + (46\beta_1 - 7\beta_2) \bar{A}^2) H^3 + 576 \beta_4^2 \bar{A}^2 H^5 + 4 \bar{A} (\beta_1 + 4\beta_4 H^2)(\beta_1 + 12\beta_4 H^2) \dot{\bar{A}} \Big) \dot{H} \nonumber\\
  && + 2 \bar{A}^2 (\beta_1 + 12\beta_4 H^2)^2 \dot{H}^2 \Big) + 4 G \bar{J} \pi H^2 (16\beta_4 H^4 + 2\beta_1 \dot{H} + H^2 (4\beta_1 - \beta_2 + 16\beta_4 \dot{H})) \bar{\rho}_{m,n} \Big] \vec{k}^2 \nonumber\\
  &&  -216 G \bar{J} \pi \bar{A} a^4 \big(16\beta_4 H^5 + 2\beta_1 H \dot{H} + H^3 (4\beta_1 - \beta_2 + 16\beta_4 \dot{H})\big)^2 \bar{\rho}_{m,n},\label{F1-20265111942}
\end{eqnarray}
\begin{eqnarray}
  F_{3}(t,\vec{k})&=& \bar{A}^2 a (\beta_1 + 4\beta_4 H^2)\left(3 H (\beta_1 + 4\beta_4 H^2) (\bar{A} H + \dot{\bar{A}}) + \bar{A}(\beta_1 + 12\beta_4 H^2) \dot{H} \right)^2 \vec{k}^4  -6 G \bar{J} \pi H^2 \bar{\rho}_{m,n}(\beta_1 \nonumber\\
  && + 4\beta_4 H^2) \Big[ H \big( H (-2 + \bar{A}^2 (8\beta_1 + \beta_2 + 24\beta_4 H^2)) + 8 \bar{A}(\beta_1 + 6\beta_4 H^2) \dot{\bar{A}} \big) +2 \bar{A}^2 (\beta_1 + 12\beta_4 H^2) \dot{H} \Big] \vec{k}^2,\label{F2-20265111942}
\end{eqnarray}
\begin{eqnarray}
  F_{4}(t,\vec{k})&=&\vec{k}^4 \bar{A}^2 a(\beta_1 + 4\beta_4 H^2) \Big[ 2 H^2(\beta_1 + 4\beta_4 H^2)\big( \vec{k}^2 + 3 a^2 ( 9\beta_1 \dot{\bar{A}}^2 + H^2 (-2 + 36\beta_4 \dot{\bar{A}}^2) - 2 \dot{H} ) \big) \nonumber\\
  && + \bar{A}^2 \Big(H^2 (\beta_1 + 4\beta_4 H^2) \big( \vec{k}^2 (4\beta_1 - \beta_2) - 6 H^2 ( -4 \vec{k}^2 \beta_4 + a^2 (\beta_1 - 4\beta_2 + 12\beta_4 H^2)) \big) \nonumber\\
  && + 2 \Big( \vec{k}^2 \beta_1^2 + 2\beta_1 (8 \vec{k}^2 \beta_4 + 3 (-\beta_1 + \beta_2) a^2) H^2 + 48\beta_4 (\vec{k}^2 \beta_4 + (-2\beta_1 + \beta_2) a^2) H^4 \nonumber\\
  && - 288\beta_4^2 a^2 H^6 \Big) \dot{H} - 6 a^2 (\beta_1 + 4\beta_4 H^2) (\beta_1 + 12\beta_4 H^2) \dot{H}^2 \Big) + 2 \bar{A} H (\beta_1 + 4\beta_4 H^2) \dot{\bar{A}} \Big( 2 \vec{k}^2 \beta_1 \nonumber\\
  && + 3 a^2 \big( 48\beta_4 H^4 + 2\beta_1 \dot{H} + H^2 (8\beta_1 + 3\beta_2 + 48\beta_4 \dot{H}) \big) \Big)\Big] +12 G \bar{J} \vec{k}^2 \pi H^2 (\beta_1 + 4\beta_4 H^2) \nonumber\\
  && \Big(6 a^2 H^2 - 24 \bar{A} a^2 H (\beta_1 + 6\beta_4 H^2) \dot{\bar{A}} + \bar{A}^2 \big( -2 \vec{k}^2 \beta_1 + H^2 ( -8 \vec{k}^2 \beta_4 + a^2 (-9\beta_2 + 24\beta_4 H^2) ) \nonumber\\
  && + 6 a^2 (\beta_1 + 4\beta_4 H^2) \dot{H} \big)\Big) \bar{\rho}_{m,n},\label{F3-20265111942}
\end{eqnarray}
\begin{eqnarray}
  F_{5}(t,\vec{k})&=&\Big[ 12 G \vec{k}^2 \pi a^2 H^2 (\beta_1 + 4\beta_4 H^2) 
  \big( 16 \beta_4 \bar{A}^2 H^3 + 2\beta_1 \bar{A} \dot{\bar{A}} + 24\beta_4 \bar{A} H^2 \dot{\bar{A}} + H ( -2 + \bar{A}^2 (6\beta_1 - \beta_2 \nonumber\\
  && + 16\beta_4 \dot{H}) ) \big) - 72 G \pi a^4 H^2 \Big( 32 \beta_4^2 \bar{A}^2 H^7 + 4(-4\beta_1 + 3\beta_2)\beta_4 \bar{A} H^4 \dot{\bar{A}} + 2\beta_1^2 \bar{A} \dot{\bar{A}} \dot{H} + \beta_1 \bar{A} H^2 \dot{\bar{A}} ( -4\beta_1 \nonumber\\
  && + 3\beta_2 + 8\beta_4 \dot{H} ) + 2\beta_1 H \dot{H} ( -1 + 2 \bar{A}^2 (\beta_1 + 2\beta_4 \dot{H}) ) + 8\beta_4 H^5 ( -1 + \bar{A}^2 (3\beta_1 - \beta_2 + 8\beta_4 \dot{H}) ) \nonumber\\
  && + H^3 \big( -2(\beta_1 + 4\beta_4 \dot{H}) + \bar{A}^2 \big( 4\beta_1^2 - 2\beta_1\beta_2 + \beta_2^2 + 8\beta_4 \dot{H} (4\beta_1 - \beta_2 + 4\beta_4 \dot{H}) \big) \big) \Big) \Big] \bar{\rho}_{m,n},\label{F4-20265111942}
\end{eqnarray}
\begin{eqnarray}
  F_{6}(t,\vec{k})&=&\Big[ 2 G \vec{k}^2 \pi H^2 (\beta_1 + 4\beta_4 H^2) \big( (-2 + (2\beta_1 + \beta_2) \bar{A}^2) H + 2 \bar{A}(\beta_1 + 12\beta_4 H^2) \dot{\bar{A}} \big) \nonumber\\
  && + 12 G \pi a^2 H^2 \Big( H^2 (\beta_1 + 4\beta_4 H^2) \big( 2 H (1 + 2 \bar{A}^2 (\beta_1 - \beta_2 + 6\beta_4 H^2)) + (4\beta_1 - 3\beta_2) \bar{A} \dot{\bar{A}} \big) \nonumber\\
  && - 2 \Big( \beta_1 (-1 + \beta_2 \bar{A}^2) H  - 4\beta_4 (1 + (6\beta_1 - 2\beta_2) \bar{A}^2) H^3 - 96 \beta_4^2 \bar{A}^2 H^5 + \beta_1 \bar{A} (\beta_1 \nonumber\\
  && + 4\beta_4 H^2) \dot{\bar{A}} \Big) \dot{H} + 8 \beta_4 \bar{A}^2 H (\beta_1 + 12\beta_4 H^2) \dot{H}^2 \Big) \Big]\bar{\rho}_{m,n}.\label{F5-20265111942}
\end{eqnarray}
}
\par
{The specific form of $F_{7}$ in the small-scale-limit approximate Lagrangian~\eqref{Lsk-20265112138} is
\begin{eqnarray}
  F_{7}(t)&=& -\frac{8 \pi G \bar{J} \bar{\rho}_{m,n}}{a^2} + a \Big( 9216 \beta_4^3 H^7 \bar{A}^3 \dot{\bar{A}} - 8 \beta 1 H\bar{A}^3 \dot{\bar{A}} (1 - 2 \beta_1) (\beta_1 -  \beta_2 - 8 \beta_4 \dot{H}) - 128 \beta_4^2 H^5 \bar{A}^3 \dot{\bar{A}} \big(5  \nonumber\\
    && - 38 \beta_1 + 10 \beta_2 + 48 \beta_4 \dot{H}\big) - 64 \beta_4 H^3 \bar{A} \dot{\bar{A}}\big(1 + \bar{A}^2 (- \beta_2 + \beta_1 (2 - 11 \beta_1 + 6 \beta_2) + 4 (1 - 4 \beta_1) \beta_4\dot{H})\big) \nonumber\\
    && + 4 \beta_1 \bar{A}^2(1 - 2 \beta 1) \big(2 \beta_1 \dot{\bar{A}}^2 - \dot{H} (2 - (2 \beta 1 + \beta 2) \bar{A}^2) + 2 \beta_1 \bar{A} \ddot{\bar{A}}\big) - 128 \beta_4^2 H^6 \bar{A}^2 \big(6 - (2 + 6 \beta_1 - 3 \beta_2) \bar{A}^2 \nonumber\\
    && + 24 \beta_4 \dot{\bar{A}}^2 +24 \beta_4 \bar{A} \ddot{\bar{A}}\big) + 16 \beta_4 H^4 \bar{A}^2 \big(2 - 24 \beta_1 + 8 \beta_2 + 8 \beta_4 \big((5 - 14 \beta_1) \dot{\bar{A}}^2 - 6 \dot{H}\big) \nonumber\\
    && + \bar{A}^2 \big( (2 \beta_1 -\beta_2)(3 + 12 \beta_1 - 4 \beta_2) + 8\beta_4 \dot{H} (4 + 6 \beta_1 - 5 \beta_2) \big) + 8 (3 - 14 \beta_1) \beta_4 \bar{A} \ddot{\bar{A}}\big) \nonumber\\
    && + H^2 \big( 4 - 4 \bar{A}^2 \big( 12 \beta_1^2 + \beta_2 - 8 \beta_1 \beta_2 - 8 \beta_4 (2 \beta_1 (2 - 5 \beta_1) \dot{\bar{A}}^2 + (1 - 4 \beta_1) \dot{H})\big) \nonumber\\
    && + \bar{A}^4 \big((2 \beta_1 -  \beta_2) (2 \beta_1 (3 + 12 \beta_1 - 8 \beta_2) -  \beta_2) + 16 \beta_4 \dot{H} (2 \beta_1 (3 + 4 \beta_1 - 6 \beta_2) -  \beta_2 + 16 \beta_4 \dot{H})\big) \nonumber\\
    && + 64 (2 - 5 \beta_1) \beta_1 \beta_4 \bar{A}^3 \ddot{\bar{A}} \big) \Big) \Big/\big(4 \bar{A}^2 (\beta_1 + 4 \beta_4 H^2) (1 - 2 \beta_1 - 8 \beta_4 H^2) \big). \label{F8-20265112141}
\end{eqnarray}
}

\bibliographystyle{unsrt}
\bibliography{referenceData}

\end{document}